\documentclass[aps,tightenlines,floats,twocolumn,prd,superscriptaddress,floatfix,showpacs]{revtex4-1}
\usepackage{graphicx}
\usepackage{epsf}
\usepackage{amsfonts}
\usepackage{amsmath}
\usepackage{amssymb}
\usepackage{bbold}
\usepackage{bm}
\usepackage{bbm}
\usepackage{slashed}
\usepackage{dcolumn}
\usepackage{rotating,array}
\makeatletter
\DeclareMathOperator{\intlh}{\mathop{=\hspace{-3.9mm}\int}}
\DeclareMathOperator{\intc}{\mathop{-\hspace{-3.0mm}\int_{-1}^{1}}\hskip 2pt}
\DeclareMathOperator{\intlc}{\mathop{-\hspace{-2.6mm}\int}}
\DeclareMathOperator{\inth}{\int_{-1}^{1}\mkern-30.6mu =
\hskip 2pt  {}}
 \makeatother

\begin{document}

\title{High-precision methods for Coulomb, linear confinement and Cornel potentials in momentum
space}

\author{V. V. Andreev}
\affiliation{Department of Physics and Information Technology, Francisk Skorina Gomel State University, Gomel,
Belarus,\\
Samara University, Samara, Russia}

\date{\today}

\begin{abstract}
We use special quadrature formulas for singular and hypersingular integral to numerically
solve the Schr\"{o}dinger equation in momentum space with the linear confinement
potential, Coulomb and Cornell potentials. It is shown that the eigenvalues of the
equation can be calculated with high accuracy, far exceeding other calculation methods.
Special methods of solution for states with zero orbital angular momentum are considered.
\end{abstract}


\maketitle

\section{Introduction}

A numerical study of some relativistic  QCD-motivated models is reduced to solving the
problems in momentum space (for instance, Bethe–Salpeter equation \cite{Bete:1951},
spinless Salpeter equation \cite{Salpeter:1952}, CST model \cite{Savkli:1999me},
Poincar\'{e}-invariant quantum mechanics (or relativistic Hamiltonian dynamics) approach
for description of bound states \cite{Keister:1991sb} and others. Typically  these
integral equations are an integral equations and reduced to the Schr\"{o}dinger equation
in the nonrelativistic limit.

Advantages of using momentum representation for solving physics problems have long
attracted the attention of researchers for a long time\cite{Salpeter:1957,EyrD:1986re}.
In momentum space, in contrast to coordinate space, relativistic effects are much
simpler. For example  there is no need for additional constructions related to the
definition of the relativistic kinetic energy operator
$T(k)=\sqrt{k^2+m_1^2}+\sqrt{k^2+m_2^2}-m_1-m_2$.

The momentum space code has an additional advantage of being easily adaptable to
relativistic equations. It is also relatively easy to obtain a relativistic interaction
potential with the use of appropriate elastic scattering amplitudes \cite{Lucha:1991jy},
since the calculation is carried out initially in momentum space, which here arises
naturally. In momentum space formulation is also flexible means of incorporating such
dynamical effects as finite size of particles, vacuum polarization and so on.

However, the problem of using momentum space is aggravated by the fact that even the
simplest interaction potentials in the momentum representation lead to integrals with
singularities.

At present, there are many papers devoted to the solution of integral equations for bound
states with singular kernels. So in the Refs.
\cite{gammel1973bethe,PhysRevC.18.932,mainland2001logarithmic,PhysRevA.50.2075,PhysRevA.50.3609,Maung:1993vg,Chen2013}
various methods of numerical solution of equations with a logarithmic singularity are
developed.

Equations with linear confinement potentials containing a double-pole singularity are
considered in Refs.
\cite{EyrD:1986re,SpencD:1993tb,PhysRevD.47.3027,Norbury:1992jv,Tang:2001ii,Deloff:2006xx,PhysRevD.88.076006,2012PhRvD..86c6013C,Leitao:2014jha}.
The subtraction technique (Land\'{e}-subtracted approach) that isolates the singularity
in an integral that can be evaluated analytically is most often used.

Therefore, the accuracy of solutions for a number of problems with Coulomb and linear
confining potentials was relatively low ($10^{-4}\div
10^{-6}$)~\cite{Norbury:1992jv,PhysRevA.50.2075,Tang:2001ii,Deloff:2006xx}, though it is
possible to reach a higher accuracy in coordinate space  $\sim 10^{-11} \hm \div
10^{-13}$ \cite{Kang:2006jd}.

The problem of accuracy in calculating characteristics of the bound quantum systems has
more than just an academic nature. A high precise calculation of various energy
corrections of the hydrogen-like systems is an relevant problem since the experimental
measurements of such values are performed with high accuracy  $\sim 10^{-13}$
\cite{PhysRevLett.79.2646,Liu:1999iz}.

Thus, when calculating characteristics of the bound quantum systems, one should allocate
the problem of developing computational methods and the development of mathematical
methods, which would allow one to simplify the calculation schemes and obtain results
with a high degree of accuracy required for the experiment.

The most promising method to increase the accuracy of solution of integral equations of
bound systems with singular kernels, is the method of quadratures, where the weight
factors depend upon the location of the singularity.

The idea of inclusion of singularities into the weight factors is not new and it is
actively used in the numerical calculations of singular integrals \cite[et
al.]{Chan2003683,bichi2014,Chen2011636,Sheshko1976en}. In \cite{Deloff:2006hd}, such an
approach was used in solving Schr\"{o}dinger equation  with the Coulomb potential
(logarithmic singularity), which allowed one to increase an accuracy of solution up to
$\sim 10^{-13} \div 10^{-14}$.

The aim of this work is to develop methods for the precision calculation of energy
spectra of the Schr\"{o}dinger equation in the momentum representation with Coulomb,
linear confining, and Cornell potentials.

This paper is organized as follows. In the Sec.~\ref{sect-1} we review the elements of
the method for solving integral equations using quadrature formulas. In the
Sec.~\ref{sect-2}  we present Land\'{e}-subtracted method of solving equations with
singular kernels.

In Sec.~\ref{sect-3} quadrature scheme for the numerical calculation of singular
integrals are presented. In the section, a method  for obtaining quadrature formulas for
integrals in which singularities are included in weight factors is given. The weight
factors are calculated analytically for various versions of the integrands using
Chebyshev polynomials in Sec.~\ref{sect-4}.

In Sec.~\ref{sect-5}, \ref{sect-6} and \ref{sect-7} we review the numerical results of
the Coulomb, linear confinement and Cornel potentials in momentum space with the help of
quadrature rules that were obtained in Sec.~\ref{sect-4}. In Sec.~\ref{sect-8} we
summarize our results and draw our conclusions.

A subtraction term for the logarithmic singularity is given in Appendix \ref{pril-a}. The
Appendix \ref{pril-b} contains a brief information about the Chebyshev polynomials needed
for calculating quadrature formulas.

The formulas needed to solve the Schr\"{o}dinger equation in the coordinate space by the
variational method are presented in Appendix \ref{pril-3}. Hereafter $\hbar=c=1$ units
are adopted.

\section{The method of solving integral equations}
\label{sect-1}

In coordinate space the quantum system with reduced mass $\mu$ and the binding energy $E$
obeys the Schr\"{o}dinger equation
\begin{equation}
-{\frac{\nabla^{2}}{2\mu} }\psi\left(\mathbf{r}\right) +
\widetilde{V}\left(\mathbf{r}\right)\psi\left(\mathbf{r}\right) =
E\,\psi\left(\mathbf{r}\right) \; .\label{wfeq2}
\end{equation}

Applying the Fourier transformations according to
\begin{eqnarray}
\phi\left(\mathbf{k}\right)&=&\int\psi\left(\mathbf{r}\right) e^{i \mathbf{k} \cdot
\mathbf{r}} \mathrm{d} \mathbf{r}\; , \label{wfft1}
\\
\psi\left(\mathbf{r}\right) &=& \int \phi\left(\mathbf{k}\right) e^{-i \mathbf{k} \cdot
\mathbf{r}} \mathrm{d}\mathbf{k} \; ,\label{wfft2}
\end{eqnarray}
one obtains a Eq.~(\ref{wfeq2}) in momentum space
\begin{equation}
\frac{\mathbf{k}^2}{2 \hskip 1pt \mu}\phi\left(\mathbf{k}\right) + \int
V\left(\mathbf{k}-\mathbf{k}^{\prime}\right)\phi\left(\mathbf{k}^{\prime}\right)\,\mathrm{d}\mathbf{k}^{\prime}
= E\,\phi\left(\mathbf{k}\right)\; . \label{wfeq1}
\end{equation}

The Schr\"{o}dinger equation (\ref{wfeq1}) for centrally symmetric potentials
$\widetilde{V}(\left|\mathbf{r}\right|)=\widetilde{V}\left(r\right)$ after partial
expansion can be written as follows:
\begin{multline}
\frac{k^2}{2 \hskip 1pt \mu}\phi_{n
\ell}(k)+\int_{0}^{\infty}V_{\ell}(k,k^{\prime})\phi_{n \ell}(k^{\prime})k^{\prime
2} \mathrm{d}k^{\prime}\\
= E_{n \ell}\phi_{n\hskip 1pt \ell}(k) \;, \; \;
k=\left|\mathbf{k}\right| \; ,\label{eq1}
\end{multline}
where wave function $\phi_{n\ell}(k)$ is the radial part of $\phi\left(\mathbf{k}\right)$
and $V_\ell(k,k^{\prime})$ denotes the $\ell$-th partial wave projection of the centrally
symmetric potential
\begin{equation}
V_\ell(k,k^{\prime})= \dfrac{2}{\pi} \int_{0}^{\infty} j_\ell\left(k^{\prime} \hskip 1pt
r\right)  j_\ell \left(k \hskip 1pt r\right) \widetilde{V}\left(r\right) \hskip 1pt  r^2
\mathrm{d}r, \label{prel3}
\end{equation}
where $j_\ell\left(x\right)$ is the spherical Bessel function .

The numerical solution of integral equation (\ref{eq1}) will be turned into a finite
matrix equation with help of the quadrature formulas for the integrals in this equation.
At the first stage we make the transition from the  semi-infinite interval of integration
 $\left(0, \infty \right[$ to the ``standard'' interval  $[-1,1]$ by means of the change of variables
\begin{equation}
\int_0^\infty f\left({k}\right) \mathrm{d}{k} = \int_{-1}^{1}
f\left({k}({t})\right)\frac{\mathrm{d}{k}}{\mathrm{d}{t}} \mathrm{d} {t} \; .\label{zam}
\end{equation}
The function  ${k\left(t\right)}$ satisfies the boundary conditions
\begin{equation}
{k\left(t=-1\right)}=0\; ,\; \; \; {k\left(t=1 \right)}=\infty \;.
\label{gran}
\end{equation}
Among various possibilities, the following mappings of the domain $\left(0, \infty
\right)$ onto $(-1,1)$ are used more frequently
\cite{Bielefeld:1999fj,Savkli:1999me,Tang:2001ii,Iersel2000,Deloff:2006xx}:
\begin{eqnarray}
&&  k(t) = \beta_{0} \hskip 1pt \frac{1+t}{1-t}\;,\label{an3} \\
&&  k(t) = \beta_{0} \hskip 2pt \hskip 1pt \sqrt{\frac{1+t}{1-t}} \label{an3x}
\end{eqnarray}
or
\begin{eqnarray}
&& k(t)= -\beta_{0}\log\left|\frac{1-t}{2}\right|\; ,\nonumber
\\
&& k(t)= \beta_{0} \tan\left[\frac{\pi}{4}\left(1+t\right)\right] \;,\label{mapfunction2}
\end{eqnarray}
where $\beta_{0}$ is a  numeric parameter. It can be used for the additional control of
the convergence rate of numerical process.

The standard approach is based on the approximation of integral (\ref{zam}) by means of
the quadrature formula
\begin{equation}
\int_0^\infty f\left({k}\right) \mathrm{d} {k} \approx \sum_{j=1}^{N} \tilde{\omega}
_{j}\hskip 2pt f\left({k}_{j}\right)\; , \label{zamg}
\end{equation}
where $N$ is the number of abscissas and the $\tilde{\omega} _{j}$ are related to the
tabulated $\omega_{j}$ weight factors for the interval $(-1,1)$  by the relationship:
$\tilde{\omega} _{j}=({dk}/\mathrm{d}{t})_{j}\; \omega_{j}\; $.

Now that we have approximated the integral by this finite sum, we can take the momentum
variable $k$ equal to the abscissas. As a result, the numerical solution of integral
equation (\ref{eq1}) can be reduced to the eigenvalue problem for the matrix $H$ which
arises when using the quadrature formulas of type (\ref{zamg}) for the integrals:
\begin{equation}
\sum\limits_{j=1}^{N}\hskip 1pt H\left(k_i, \hskip 1pt k_j\right) \phi(k_{j}) =
\sum\limits_{j=1}^{N}\hskip 1pt H_{i j} \phi_{j}= E^{(N)} \phi_i\; , \label{hij}
\end{equation}
where $E^{(N)} \approx  E_{n \ell}$ and the matrix-elements $H_{i j}$ are given by:
\begin{equation}
H_{i j} = \frac{k_{j}^{2}}{2\hskip 2pt \mu}\delta _{i,j} + \tilde{w}_{j}\hskip 1pt
k^2_j\hskip 1pt {V}_{l}(k_{i},k_{j})\; . \label{hijz}
\end{equation}

However, the description of bound states in momentum space has a singular kernel for both
the Coulomb and  linear confinement potentials. Let us illustrate this statement.

The Coulomb potential
\begin{equation}
\widetilde{V}(r)=-\frac{\alpha}{r} \;  \label{eqh2}
\end{equation}
in momentum space has the form
\begin{equation}
V_{\ell}(k,k^{\prime})=-\hskip 2pt \frac{\alpha Q_{\ell}(y)}{\pi(k k^{\prime})}\; ,
\label{eqh3}
\end{equation}
where the coupling parameter $\alpha$ is dimensionless.

Parameter $y$ in (\ref{eqh3}) is the combination of momenta
\begin{equation}
y=\frac{k^2+k^{{\prime}^2}}{2 k\hskip 1pt  k^{\prime}}\;  , \label{eq4}
\end{equation}
and the $Q_\ell(y)$ is Legendre polynomial of the second kind:
\begin{equation}
Q_{\ell}(y)=P_{\ell}(y)Q_0(y) - w_{l-1}(y)\; , \label{eq5}
\end{equation}
\begin{multline}
Q_0(y)=\frac{1}{2}\log\left|\frac{1+y}{1-y}\right|\; ,\\
w_{l-1}(y)=\sum_{n=1}^{l}\frac{1}{n}P_{n-1}(y)P_{l-n}(y)\; . \label{eq6}
\end{multline}
In Eq.~ (\ref{eq5}) $P_{\ell}(y)$ is the Legendre polynomial of the first kind.

From (\ref{eq5}) and (\ref{eq6})  it follows that potential (\ref{eqh3}) has a
logarithmic singularity in the case where $k=k^{\prime}\left(y=1\right)$.

The linear confinement potential with parameter $\sigma$
\begin{equation}
V(r)=\sigma r\; , \label{eq2}
\end{equation}
in momentum space is written in the form
\begin{equation}
V_{\ell}(k,k^{\prime})=\frac{\sigma Q^{\prime}_{\ell}(y)}{\pi(k \hskip 1pt
k^{\prime})^2}\; . \label{eq3}
\end{equation}

With the help of (\ref{eq5}) and (\ref{eq6}) we find that the derivative
$Q^{\prime}_{\ell}(y)$ in Eq.~ (\ref{eq3}) is given by the relation
\begin{equation}
Q^{\prime}_{\ell}(y)=P^{\prime}_{\ell}(y)Q_0(y)+P_{\ell}(y)Q^{\prime}_0(y)-w^{\prime}_{l-1}(y)\;
, \label{eq7}
\end{equation}
\begin{equation}
Q^{\prime}_0(y)=\frac{1}{1-y^2}=-\left(\frac{2k k^{\prime}}{k^{\prime}+k}\right)^2
\frac{1}{(k^{\prime}-k)^2}\; . \label{eq8}
\end{equation}
As follows from (\ref{eq8}), the function  $Q^{\prime}_{\ell}(y)$ is hypersingular in the
case $k=k^{\prime}$, and  $V_{\ell}(k,k^{\prime})$ consequently the potential itself is
also hypersingular.

As follows from the above, the problem of calculating the elements (\ref{hijz}) for the
Coulomb and linear confinement  potentials is not complex if $i \neq j$. However, for
$i=j \; \left({k}={k^{\prime}}\right)$ it is not possible to directly compute $H_{i j}$
due to the presence of singularities.

\section{Land\'{e}-subtracted method of solving equations
with singular kernels} \label{sect-2}

\subsection{Linear confinement  potential}

Consider some methods of solving the Schr\"{o}dinger equation with a hypersingular
kernel, using as an example the equation (\ref{eq1})  with a linear potential (\ref{eq3})
for $\ell=0$, namely,
\begin{multline}
\left(E_{n 0} - \frac{k^2}{2\mu} \right)\phi _{n \hskip 1pt 0}(k)
\\= \frac{\sigma}{{\pi
{k^2}}}\int_0^\infty   \, Q'_0(y)\phi _{n \hskip 1pt 0} (k^{\prime})\mathrm{d}k^{\prime}
\; .\label{pr1}
\end{multline}
Having substituted (\ref{eq8}) into (\ref{pr1}, one obtains explicitly the equation with
hypersingular kernel
\begin{multline}
\left(E_{n 0} - \frac{k^2}{2\mu} \right)\phi _{n \hskip 1pt 0}(k) \\
= -\frac{4 \sigma}{{\pi }}\int_0^\infty \,  \left(\frac{
k^{\prime}}{k^{\prime}+k}\right)^2 \frac{1}{(k^{\prime}-k)^2} \phi _{n \hskip 1pt 0}
(k^{\prime})\mathrm{d}k^{\prime} \; .\label{pr2}
\end{multline}

To obtain the solution, the most frequently used method assumes the ``reduction'' of the
singularity $\sim 1/(k-k^{\prime})^{2}$ with the help of a counter term (the Land\'{e}
subtraction method)
\cite{Kahana:1993yd,Norbury:1992jv,PhysRevD.47.3027,Tang:2001ii,2012PhRvD..86c6013C,Leitao:2014jha}:
\begin{equation}
\int_0^\infty \,dk\;{Q'_0}\left( y \right) = 0\;. \label{contr1}
\end{equation}

Then Eq.~ (\ref{pr2}) after adding (\ref{contr1}) is written in the form
\begin{multline}
\left(E_{n 0} - \frac{k^2}{2\mu} \right)\phi _{n \hskip 1pt 0}(k)= -\frac{4
\sigma}{{\pi }} \\
\times\int_0^\infty \,  \left(\frac{ k^{\prime}}{k^{\prime}+k}\right)^2
\frac{1}{(k^{\prime}-k)^2} \left(\phi _{n \hskip 1pt 0} (k^{\prime})-\phi _{n \hskip 1pt
0} (k)\right)\mathrm{d}k^{\prime} \; .\label{pr3v}
\end{multline}

For calculating the energy spectrum, the function $\phi(k)$ can be expanded in a complete
set of functions, for example \cite{Kahana:1993yd,Tang:2001ii},
\begin{multline}
g_i^{A}(k)=\frac{1}{\left(i/N\right)^{2}+k^4}\; ,\;  g_i^{B}(k)=\exp\left[-\hskip 1pt
\frac{i^2 k^2}{N}\right]\; ,\label{pr4}
\end{multline}
where $N$ is the maximum number of basis functions used in the expansion and $i=1,\ldots
N$.

In another method, quadrature formulas like (\ref{zamg}) are used. In \cite{Tang:2001ii},
to improve the efficiency of the solution, correction method was proposed that includes
additional processing of the hypersingular term to fully reduce the singularity. As a
result, the authors of \cite{Tang:2001ii}  managed to increase the accuracy of energy
spectrum from  $10^{-3}$ (see \cite{Kahana:1993yd,Tang:2001ii}) to $10^{-6}$ for the
ground state ($n=1$) up to $10^{-2} \div 10^{-3}$ for $n=2,3,4$ at $N=1400$.

In \cite{Deloff:2006xx}, the correction method  was criticized, and to improve an
accuracy it was proposed to transform integral equation (\ref{pr2}) into an
integral-differential  equation by means of the relation
\begin{multline}
\int_0^\infty \frac{f(k,k^{\prime})}{(k^{\prime}-k)^2}\phi _{n \hskip 1pt
0}(k)\mathrm{d}k^{\prime}\\
=\int_0^\infty \frac{\mathrm{d}k^{\prime}}{(k^{\prime}-k)}\frac{\partial}{\partial
k^{\prime}}\left[f(k,k^{\prime})\phi _{n \hskip 1pt 0}(k)\right]\; ,\label{del1}
\end{multline}
obtained after integration by parts. In this case, the function $f(k,k^{\prime})$ does
not contain the singularities at $k=k^{\prime}$.

After using (\ref{del1}), Eq.~ (\ref{pr2}) takes the form
\begin{multline}
\left(E_{n 0} - \frac{k^2}{2\mu} \right)\phi _{n \hskip 1pt 0}(k) =  -\frac{4 \sigma}{{\pi }}
\\ \times \int_0^\infty
\,\frac{\mathrm{d}k^{\prime}}{(k^{\prime}-k)}\left\{\frac{\partial \phi _{n \hskip 1pt 0}
(k^{\prime})}{\partial k^{\prime}}+\phi _{n \hskip 1pt 0}
(k^{\prime})\frac{\partial}{\partial
k^{\prime}}\right\}\left(\frac{k^{\prime}}{k^{\prime}+k}\right)^{2}\; ,\label{pr5}
\end{multline}
in this case, to calculate the derivative of the unknown function $\phi _{n 0}
(k^{\prime})$, the following relationship was used after performing discretization with
the help of quadrature formulas
\begin{equation}
\left\{\frac{\partial \phi (k)}{\partial k}\right\}_{k=k_i}=\sum\limits_{j=1}^{N}D_{ij}\phi (k_j)
\; .\label{pr6}
\end{equation}
The matrix elements $D_{ij}$  can be calculated after expanding the function by means of
the interpolation polynomial $ G_i\left(t\right)$
\begin{equation}
\phi(k) \approx \sum\limits_{i= 1}^N G_i\left(k\right)\hskip 1pt \phi\left( {k _{i}} \right) \;
.\label{eqint10s}
\end{equation}
This method made it possible to obtain eigenvalues $E$  with the accuracy at $\sim
10^{-6}$ for the ground state and radially excited states ($n=1,\ldots 4$ ).

\subsection{Coulomb potential}
Consider the methods of solving Eq.~ (\ref{eq1}) with Coulomb potential (\ref{eqh3}),
i.e., with the kernel having a logarithmic singularity
\begin{equation}
\left(E_{n \ell} - \frac{k^2}{2\mu} \right)\phi _{n \ell}(k) = -\frac{\alpha}{{\pi {k
}}}\int_0^\infty   \, Q_l(y)\phi_{n
 \ell} (k^{\prime})k^{\prime} \mathrm{d}k^{\prime} \;
.\label{prh1}
\end{equation}

With the Eqs.~(\ref{eq5}) and (\ref{eq6}), the relationship (\ref{prh1}) is transformed
as
\begin{eqnarray}
&& \left(E_{n  \ell} - \frac{k^2}{2\mu} \right)\phi _{n \hskip 1pt \ell}(k) =
-\frac{\alpha}{{\pi\hskip 1pt k }}\int_0^\infty \, P_{\ell}(y) Q_0(y) k^{\prime}
\phi_{n \hskip 1pt \ell} (k^{\prime})\mathrm{d}k^{\prime} \nonumber \\
&& +\frac{\alpha}{{\pi\hskip 1pt  k }} \int_0^\infty \, w_{\ell-1}(y)  k^{\prime} \phi
_{n \hskip 1pt \ell} (k^{\prime})\mathrm{d}k^{\prime}\; .\label{prh2}
\end{eqnarray}
A number of methods have been developed to handle the problems associated with the
logarithmic singularity. Many of them use a Land\'{e} subtraction technique  to isolate
the singularity by means of identity \cite{Tang:2001ii,PhysRevA.50.2075,Leitao:2014jha}:
\begin{equation}
\int_0^\infty \,dk\;\frac{{Q_0}\left( y \right)}{k} = \frac{\pi^{2}}{2}\;
.\label{contrh1}
\end{equation}
The Land\'{e}-subtraced  equation (\ref{prh2}) with a Coulomb potential  is written in
the form
\begin{eqnarray}
&&\left(E_{n  \ell} - \frac{k^2}{2\mu} + \alpha \frac{\pi}{2}\hskip 1pt k\right)\phi _{n
\hskip 1pt \ell}(k) \nonumber \\
&&= -\frac{\alpha}{\pi}\int_{0}^{\infty} \, P_{\ell}(y) Q_0(y) \left[\frac{k^{\prime}}{{k
}}\phi_{n \hskip 1pt \ell} (k^{\prime})-\frac{k}{{k}^{\prime}}\phi_{n \hskip 1pt \ell}
(k)\right]
\mathrm{d}k^{\prime} \nonumber \\
&& +\frac{\alpha}{{\pi\hskip 1pt  k}} \int_0^\infty \, w_{\ell-1}(y)  k^{\prime} \phi _{n
\hskip 1pt \ell} (k^{\prime})\mathrm{d}k^{\prime} \;. \label{LandeC}
\end{eqnarray}
Since the bracketed term in (\ref{LandeC}) vanishes for $k =k^{\prime}$, this integral
equation can be converted to a matrix equation without singularity.

However, counter term (\ref{contrh1}) is not very convenient in solving Eq.~ (\ref{prh1})
for $\ell\neq 0$. Due to this, in \cite{PhysRevC.18.932} it was proposed to subtract the
term (Kwon-Tabakin-Land\'{e} technique \cite{landau1983coupled})
\begin{eqnarray}
&& S_{\ell}=\int_0^\infty \,\frac{dk}{k}\;\frac{{Q_\ell}\left( y \right)}{{P_\ell}\left(
y
\right)} \nonumber \\
&&= \frac{\pi^{2}}{2} - \int_0^\infty \,\frac{dk}{k}\;\frac{w_{\ell-1}\left( y
\right)}{{P_\ell}\left( y \right)}= \frac{\pi^{2}}{2} - \mathcal{I}_{\ell}\;
.\label{contrh2}
\end{eqnarray}

In (\ref{contrh2}), it is proposed to calculate the  integral $\mathcal{I}_{\ell}$ at the
values  $\ell=0,1,2,3 \ldots 10$ frequently used in physical applications. Thus, for
example, $\mathcal{I}_{0}=0$, $\mathcal{I}_{1}=1$, $\mathcal{I}_{2}=\sqrt{3/2}$ and
$\mathcal{I}_{4}=\left(8+5\sqrt{10}\right)/18$ (see \cite{PhysRevC.18.932}). In
\cite{Ivanov2001}, it is proposed to reduce $\mathcal{I}_{\ell}$ to the integral of the
form
\begin{multline}
\mathcal{I}_{\ell}= 2\int_1^\infty \,\frac{dy}{y}\;\frac{w_{\ell-1}\left(y
\right)}{{P_\ell}\left(y \right)\sqrt{y^2-1}}\\
\sim \mathcal{I}_{\ell}^{n}=2\int_1^\infty \,\frac{dy}{y}\;\frac{y^{n}}{{P_\ell}\left(y
\right)\sqrt{y^2-1}}\; .\label{contrh3}
\end{multline}
Integral $\mathcal{I}_{\ell}^{n}$ in \cite{Ivanov2001} can be calculated by the formula
\begin{equation}
\mathcal{I}_{\ell}^{n}=
2\sum\limits_{k=1}^{\ell}\frac{\left(\xi_{k,\ell}\right)^{n}}{P_{\ell}^{\prime}(\xi_{k,\ell})}
\frac{\arccos\left(-\xi_{k,\ell}\right)}{\sqrt{1-\xi_{k,\ell}^{2}}}\; ,\label{contrh3a}
\end{equation}
where $\xi_{k,\ell}$ is the $k$th zero of the Legendre polynomial $P_{\ell}(y)$.

We propose a generalization of counter term (\ref{contrh1}) (see,also \cite{Chen2013})
which is maximally convenient for subtraction and allows one to increase the accuracy of
solution when compared to (\ref{contrh1}) and (\ref{contrh2}):
\begin{equation}
\mathcal{C}_{\ell}=\int_0^\infty \,dk\;\frac{{Q_\ell}\left( y \right)}{k} \;
.\label{contrh4}
\end{equation}

Using the integral representation for the polynomial  $Q_{\ell}\left(y\right)$, we obtain
after several transformations the analytical expression for $\mathcal{C}_{\ell}$ in the
form (see Appendix \ref{pril-a})
\begin{multline}
\mathcal{C}_{\ell}=\int_{0}^{\infty} \frac{Q_{\ell}\left(y\right)}{k}
\mathrm{d}k=\left[\cfrac{(\ell-1)\hskip 1pt !!}{\ell \hskip 1pt !!}\right]^{2}\\
\times \left\{\begin{array}{cl}
 \pi^2/2 &\;, \ell =2\hskip 1pt m \\
 2 &\;,\ell = 2\hskip 1pt m+1  \\
 \end{array}
 \right.\; ,\;  m=0,1,2\ldots
  \label{contrh5}
\end{multline}

As a result, Eq.~ (\ref{pr1}) after adding (\ref{contrh5}) is written as
\begin{eqnarray}
&& \left(E - \frac{k^2}{2\mu}+\frac{\alpha}{{\pi }}\hskip 1pt \mathcal{C}_{\ell}\hskip 1pt  k \right)\phi _{n \hskip 1pt \ell}(k) = \nonumber \\
&& = -\frac{\alpha}{{\pi }}\int_0^\infty \, Q_l(y)\left(\frac{k^{\prime}}{{k }}\phi_{n
\hskip 1pt \ell} (k^{\prime})-\frac{k}{{k}^{\prime}}\phi_{n \hskip 1pt \ell} (k)\right)
\mathrm{d}k^{\prime}\; .\label{prh3}
\end{eqnarray}

This method makes it possible to find the eigenvalues $E_{n\ell}$ with an accuracy of
$\sim 10^{-6}$ for $\ell=0$ and for $\ell=1$ with an accuracy of $\sim 10^{-5}$  for
$N=100$ the ground state ($n=1$) and radially excited states, respectively.

As one can see, the maximum possible accuracy in solving the Schr\"{o}dinger equation in
momentum space reaches $\sim 10^{-6}$ for both Coulomb and linear potential, though in
coordinate space one may reach a considerably higher accuracy $\sim 10^{-11} \hm \div
10^{-13}$ \cite{Kang:2006jd}. It is therefore necessary to find such methods of finding
the eigenvalues that are comparable with the accuracy of solutions obtained in coordinate
space.

In contrast to the Land\'{e}-subtracted approaches used in solving the Schr\"{o}dinger
equation, the main feature of the developed approach, which should increase the accuracy
of solving Eq.~ (\ref{eq1}) with singular potentials, is the inclusion of singularities
into the weight factors $\omega_{i}$ of the quadrature formula of type (\ref{zamg}).

Furthermore, we consider the general calculation method  of such weight factors using the
interpolation polynomial
\begin{equation}
G_i\left(t\right)= \frac{P_N^{\left(\alpha ,\beta\right)}\left( t\right)}{\left(t - \xi
_{i,N}\right) P_{N}^{\hskip 1pt \prime\left(\alpha,\beta\right)}\left( \xi _{i,N} \right)} \;,
\label{pr8}
\end{equation}
where $\xi _{i,N}$ are the zeroes of the Jacobi polynomial
\begin{equation}
P_N^{\left( {\alpha ,\beta } \right)} \left( {\xi _{i,N} } \right) = 0\,\,\,\left( {i =
1,2, \ldots ,N} \right)\;. \label{pr9}
\end{equation}

Of all Jacobi polynomials $P_{N}^{\left({\alpha,\beta} \right)} \left(z\right)$, it is
better to take polynomials with $\alpha,\beta = \pm 1/2$. These polynomials
$P_{N}^{\left({\pm 1/2,\pm 1/2} \right)} \left(z\right)$ associated with Chebyshev
polynomials. There are several kinds of Chebyshev polynomials. These include the
Chebyshev polynomials of the first $T_n(x)$ , second $U_n(x)$, third $V_n(x)$ and fourth
$W_n(x)$ kinds \cite{Mason2002}.

Let's introduce a function ${K}_{n}^{\left(\alpha,\beta\right)}\left(z\right)$  that
generalizes the Chebyshev polynomials by defining
\begin{equation}
{K}_{n}^{\left(\alpha,\beta\right)}\left(z\right)=\left\{
\begin{array}{cl}
T_n(z)\;,&\alpha=~~\beta=-1/2\; , \\
U_n(z)\;,&\alpha=~~\beta=~~1/2\; ,  \\
V_n(z)\;,&\alpha=-\beta=-1/2\; ,  \\
W_n(z)\;,&\alpha=-\beta=~~1/2 \; . \\
\end{array}\right. \label{w5dop}
\end{equation}

For these polynomials, the convergence of quadratures is maximal relative to other Jacobi
polynomials. Moreover, the zeroes of polynomials can be easily calculated (there are
analytical expressions) and many integrals for the weight factors with singularities are
given by relatively simple formulas
\cite{Kaya1987,Golberg1990,mason1999chebyshev,Sheshko1976en} (see Appendix \ref{pril-b}).

\section{Quadrature scheme for the evaluation of singular integrals}
\label{sect-3}

Let us find the quadrature formula for the integral
\begin{equation}
I\left(z\right)=\int_{-1}^{1} F(t) w(t) \hskip 1pt g\left(t,z\right) \mathrm{d}t
\label{eqint9a}
\end{equation}
where $g\left(t,z\right)$ is the singular function at $t=z$ and $F(t), w(t)$ are the part
of the kernel without singularities for all  $-1<t,z<1$.

For this purpose, the function $F(t)$  in  (\ref{eqint9a}) is replaced by the following
expression with the help of interpolation polynomial (\ref{pr8})
\begin{equation}
F(t) \approx \sum\limits_{i = 1}^N G_i\left(t\right)\hskip 1pt F\left( {\xi _{i,N}}
\right) \; ,\label{eqint10}
\end{equation}
where $\xi _{i,N}$  are the zeroes of Jacobi polynomial (see (\ref{pr9})).

Substituting expansion (\ref{eqint10}) into $I\left(z\right)$, the quadrature formula for
the integral takes the form
\begin{equation}
I\left(z\right) \approx  \sum\limits_{i = 1}^N \omega_{i}\left(z\right)F\left( {\xi
_{i,N} }\right)\;  \label{eqint12}
\end{equation}
with
\begin{equation}
\omega_{i}\left(z\right)= \cfrac{\widetilde{\omega}_{N}\left(z,\xi
_{i,N}\right)}{P_{N}^{\hskip 1pt \prime\left({\alpha,\beta} \right)} \left( {\xi _{i,N} }
\right)}\; ,\label{eqint13}
\end{equation}
where additional function are introduced to simplify the notation
\begin{equation}
\widetilde{\omega}_{j}\left(z,\xi\right)= \int_{-1}^1 g\left(t,z\right) \hskip 1pt
w\left(t\right) \hskip 1pt \frac{P_j^{\left(\alpha ,\beta\right)} \left(t\right)}{t -
\xi}\mathrm{d}t\; .\label{eqint13z}
\end{equation}

The use of the Christoffel-Darboux formula for the Jacobi polynomials
\begin{multline}
\sum\limits_{m = 0}^n {\frac{1} {{h_m }}} P_m^{\left( {\alpha ,\beta } \right)} \left( x
\right)P_m^{\left( {\alpha ,\beta } \right)}
\left( y \right) = \frac{{k_n }}{{k_{n + 1} h_n }}  \\
\times \frac{{P_{n + 1}^{\left( {\alpha ,\beta } \right)} \left( x \right)P_n^{\left(
{\alpha ,\beta } \right)} \left( y \right) - P_n^{\left( {\alpha ,\beta } \right)} \left(
x \right)P_{n + 1}^{\left( {\alpha ,\beta } \right)} \left( y \right)}} {{x - y}}\; ,
\label{eqint14}
\end{multline}
where
\begin{eqnarray}
k_m  &=& \frac{{\Gamma \left( {2m + \alpha  + \beta  + 1} \right)}}
{{2^m \Gamma \left( {m + \alpha  + \beta  + 1} \right)\Gamma \left( {m + 1} \right)}}\,\,, \nonumber \\
h_m  &=& \frac{{2^{\alpha  + \beta  + 1} \Gamma \left({m + \alpha  + 1} \right)\Gamma }}
{{\left({2m + \alpha + \beta  + 1} \right)\Gamma \left({m + 1}\right)
}}\nonumber \\
&\times&\frac{\left( {m +\beta  + 1} \right)}{\Gamma \left({m +\alpha+\beta +1}
\right)}\;. \label{eqint15}
\end{eqnarray}
gives the result for the weight factor in the form
\begin{equation}
\begin{gathered}
\omega_{i}\left(z\right) = \lambda_{i,N}^{(\alpha,\beta)}\sum\limits_{m = 0}^{N- 1} {
\frac{1}{{h}_m} P_m^{\left( {\alpha ,\beta } \right)} \left( {\xi _{i,N} } \right)}\hskip 1pt J_{m}^{(\alpha,\beta)}(z) \; .  \hfill \\
\end{gathered}
 \label{eqint16}
\end{equation}

The Christoffel symbols $\lambda_{m,N}^{(\alpha,\beta)}$ in (\ref{eqint16}) for the
Jacobi polynomials are defined by the relation \cite{Szego1939}
\begin{eqnarray}
\lambda_{m,N}^{(\alpha,\beta)}&=& \int_{ - 1}^1 \frac{w^{(\alpha,\beta)}(t)\hskip 1pt
P_m^{\left(\alpha,\beta\right)}\left(t\right)}{P_{N}^{\hskip 1pt
\prime\left({\alpha,\beta}
\right)} \left( {\xi_{i,N} } \right)\left(x-\xi_{i,N}\right)} \mathrm{d}t \nonumber  \\
 &=&\frac{{2^{\alpha  + \beta  + 1} \Gamma \left({N + \alpha  + 1} \right)\Gamma \left(
{N +\beta + 1} \right)}} {{\Gamma \left({N + 1}\right) \Gamma \left({N +\alpha+\beta +1}
\right)}} \nonumber
\\
&\times & \frac{1}{{\left(1-\xi_{i,N}^{2}\right)\left[P_{N}^{\hskip 1pt
\prime\left({\alpha,\beta} \right)} \left( {\xi_{i,N} } \right)\right]^{2} }}\; ,
\label{eqint17}
\end{eqnarray}
then the integral $J_{m}^{(\alpha,\beta)}(z)$ takes up the form
\begin{equation}
J_{m}^{(\alpha,\beta)}(z) =\int_{-1}^1 g\left(t,z\right){w(t)
P_m^{\left(\alpha,\beta\right)}\left(t\right) \mathrm{d}t} \; . \label{eqint20}
\end{equation}

The coefficient $\lambda_{i,N}^{(\alpha,\beta)}$ is the weight factor for the integral
$I\left(z\right)$ without the singular function $g(t, z)$, i.e.
\begin{equation}
\int_{-1}^{1} F\left(t\right)\hskip 1pt w^{(\alpha,\beta)}(t) \mathrm{d}t \approx
\sum\limits_{i =1}^N \lambda_{i,N}^{(\alpha,\beta)}F\left({\xi_{i,N}}\right) \;
,\label{eqint18}
\end{equation}
where the function $w^{(\alpha,\beta)}(t)$ is a weight function of the Jacobi polynomial
$P_{N}^{\left({\alpha,\beta} \right)} \left(x\right)$
\begin{equation}
w^{(\alpha,\beta)}(t) = \left(1-t\right)^\alpha \left(1 + t\right)^\beta \;, \; \; \;
\alpha, \beta > -1 \; .\label{eqint19}
\end{equation}

The important case of practical interest is that  in which  $w(t)=1$ and we have
\begin{equation}
\int_{-1}^1  F(t) \mathrm{d}t \approx \sum\limits_{i=1}^N \, \omega_i^{\mathrm{st}}
\hskip 1pt F\left({\xi_{i,N}}\right) \label{eqint45s}
\end{equation}
with the  weights
\begin{equation}
{\omega}_{i}^{\mathrm{st}}= \cfrac{1}{P_{N}^{\hskip 1pt \prime\left({\alpha,\beta}
\right)} \left( {\xi _{i,N} } \right)}\;\int_{-1}^1 \hskip 1pt \frac{P_i^{\left(\alpha
,\beta\right)} \left(t\right)}{t - \xi_{i,N}}\mathrm{d}t\; .\label{eqint46s}
\end{equation}
Using Eqs.~ (\ref{eqint16}) and (\ref{eqint20}) with $g(t,z)=w(t)=1$ the weights in
(\ref{eqint46s}) read
\begin{equation}
\begin{gathered}
\omega_{i}^{\mathrm{st}} = \lambda_{i,N}^{(\alpha,\beta)}\sum\limits_{m = 0}^{N- 1} {
\frac{1}{{h}_m} P_m^{\left( {\alpha ,\beta } \right)} \left( {\xi _{i,N} } \right)}  J_{m}^{\left(\alpha ,\beta\right)}\; ,  \hfill \\
\end{gathered}
 \label{eqint47s}
\end{equation}
where
\begin{eqnarray}
J_{m}^{\left(\alpha
,\beta\right)}&=&\frac{2}{m+\alpha+\beta} \nonumber \\
&\times & \left[\binom{m+\alpha}{m+1}+(-1)^m \binom{m+\beta}{m+1} \right]\; .
\label{eqint48s}
\end{eqnarray}
For example, when $\alpha=\beta=-1/2$ the relation (\ref{eqint47s}) is transformed to the
form \cite{Deloff:2006hd}
\begin{equation}
\omega_{i}^{\mathrm{st}}=-\dfrac{4}{N}\hskip 5pt  \sideset{}{ \hskip 1pt
^{\prime}}\sum\limits_{k=0}^{[(N-1)/2]}\dfrac{ T_{2 k}(\xi _{i,N})}{4k^2-1}\;
,\label{eqint49s}
\end{equation}
where the sign $\prime$ indicate that the first  term in the sum is divided by two.  The
$\left[n\right]$ symbol means that the integer part of the number $n$ is taken.

Therefore, the calculation of (\ref{eqint13}) or (\ref{eqint20}) makes it possible to
find the weight factors for quadrature formula (\ref{eqint12}) with singularities. One
important fact is the calculation of the analytical expressions, since only in this case
it is possible to increase the accuracy of calculations.

\section{Analytical expressions of weights with a singularity}
\label{sect-4}

Consider the possibility of analytical calculation of the weights for various forms of
singularities, i.e., depending on the form of the function $g\left(t,z\right)$.

\subsection{Cauchy integral}
The most known variant of (\ref{eqint9a}) in the literature is the Cauchy integral (sign
$\intlc$)
$$
g\left(t,z\right)=\frac{1}{t-z}\; , \; \; \; -1<z<1\; .
$$

There are many works for this case (see, for example \cite{Golberg1990,Sheshko1976en}),
in which the different variants of quadrature formulas are proposed. Therefore, it is
possible to obtain formulas for weight factors (\ref{eqint13}) by the direct calculation
of the integral
\begin{equation}
\widetilde{\omega}_{N}^{C}\left(z,\xi_{i,N}\right)= {\intc \frac{w\left(t\right)\hskip
1pt{P_N^{\left( {\alpha ,\beta } \right)} \left(t \right)}}{{\left(t - \xi
_{i,N}\right)\left(t-z\right) }}\mathrm{d}t}\; .\label{fz1}
\end{equation}
Using the identity
\begin{equation}
\frac{1}{\left(t - \xi
_{i,N}\right)\left(t-z\right)}=\frac{1}{z-\xi_{i,N}}\left[\frac{1}{t-z}-\frac{1}{t-\xi_{i,N}
}\right] \label{fz2}
\end{equation}
coefficients (\ref{fz1}) reduce to the form
\begin{equation}
\widetilde{\omega}_{N}^{C}\left(z,\xi_{i,N}\right)=\left\{
\begin{array}{ll}
\dfrac{\Pi_{N}^{\left(\alpha,\beta\right)}\left(z\right)-\Pi_{N}^{\left(\alpha,\beta\right)}\left(\xi
_{i,N}\right)}{\left(z-\xi_{i,N}\right)} & \;, z\neq \xi _{i,N} \;,
\hfill \\
\Pi_{N}^{\hskip 1pt \prime \left({\alpha,\beta} \right)}\left(\xi _{i,N}\right)& \;,
 z = \xi _{i,N}\;,
\hfill \\
\end{array}
\right. \label{fz3}
\end{equation}
where
\begin{equation}
\Pi_{n}^{\left(\alpha,\beta\right)}\left(z\right) =\intc w(t) \frac{P_n^{\left(\alpha
,\beta\right)}\left(t\right)}{\left(t-z\right)}\mathrm{d}t\; . \label{fz4}
\end{equation}
For calculating coefficients
\begin{equation}
\omega_{i}^{C}\left(z\right)=\cfrac{\widetilde{\omega}_{N}^{C}\left(z,\xi
_{i,N}\right)}{P_{N}^{\hskip 1pt \prime\left({\alpha,\beta} \right)} \left( {\xi _{i,N} }
\right)} \label{wc1}
\end{equation}
with high degree of accuracy, it is necessary to evaluate integral (\ref{fz4})
analytically for various forms of the function  $w(t)$.

\subsubsection{$w(t)=
w^{\left(\alpha,\beta\right)}\left(t\right)\equiv\left(1-t\right)^{\alpha}
\left(1+t\right)^{\beta}$}

The most known representation of $w(t)$ is the form with the Jacobi polynomial weight
function $P_{n}^{\left(\alpha,\beta\right)} \left(t\right)$; that is,
$$
w(t)=
w^{\left(\alpha,\beta\right)}\left(t\right)\equiv\left(1-t\right)^{\alpha}
\left(1+t\right)^{\beta}.
$$
Then for integral (\ref{fz4}) one obtains
$$
\Pi_{n}^{\left(\alpha,\beta\right)}\left(z\right)=\mathcal{\bar{Q}}_{n}^{\left(\alpha,\beta\right)}\left(z\right)\;
,
$$
where
\begin{equation}
\mathcal{\bar{Q}}_{n}^{\left(\alpha,\beta\right)}\left(z\right) =\intc
\left(1-t\right)^{\alpha} \left(1+t\right)^{\beta} \frac{P_n^{\left(\alpha
,\beta\right)}\left(t\right)}{\left(t-z\right)}\hskip 1pt \mathrm{d}t\;. \label{qn1}
\end{equation}

In the most general case at arbitrary $\alpha$ and $\beta$, the function
$\mathcal{\bar{Q}}_{n}^{\left(\alpha,\beta\right)}\left(z\right)$ is related to the
second order Jacobi polynomials $Q_{n}^{\left(\alpha,\beta\right)}\left(z\right)$ by the
relationship
\begin{equation}
\mathcal{\bar{Q}}_{n}^{\left(\alpha,\beta\right)}\left(z\right)= \left(-2\right)
(z-1)^{\alpha}\left(z+1\right)^{\beta}Q_{n}^{\left(\alpha,\beta\right)}\left(z\right)\; ,
\label{fz5}
\end{equation}
where \cite{Batemant2}
\begin{multline}
Q_{n}^{\left(\alpha,\beta\right)}\left(z\right) = 2^{\alpha +\beta +n}\hskip 2pt \frac{
\Gamma (n+\alpha +1)
\Gamma (n+\beta +1)}{\Gamma (2\hskip 1pt  n+\alpha +\beta +2)} \\
\times (z+1)^{-\beta} (z-1)^{-\alpha-n-1}\\
\times{}_2F_1\left(n+1,n+\alpha +1;2\hskip 1pt  n+\alpha +\beta +2;\frac{2}{1-z}\right) .
\label{q2r}
\end{multline}

\subsubsection{$w(t)=1$}

Consider the integral (\ref{fz4}), when $w(t)=1$. This is the most economical and natural
choice for practical calculations.

The subtraction procedure leads us to the relation
\begin{equation}
\begin{gathered}
\Pi_{n}^{\left(\alpha,\beta\right)}\left(z\right)=\mathcal{P}_{N}^{\left(\alpha,\beta\right)}\left(z\right)
= \int_{ - 1}^1 \frac{P_N^{\left(\alpha
,\beta\right)}\left(t\right)}{\left(t-z\right)}\hskip 1pt
\mathrm{d}t \hfill  \\
= \int_{ - 1}^1 \frac{P_N^{\left(\alpha ,\beta\right)}\left(t\right)-P_N^{\left(\alpha
,\beta\right)}\left(z\right)}{\left(t-z\right)}\hskip 1pt \mathrm{d}t \\+
P_N^{\left(\alpha ,\beta\right)}\left(z\right) \log\left(\frac{1-z}{1+z}\right)
\end{gathered}
\label{w0}
\end{equation}

Consider equation (\ref{w0}) when $\alpha=\beta= \pm 1/2$. Using equation (\ref{w7}), we
obtain that
\begin{equation}
\begin{gathered}
\mathcal{P}_{N}^{\left(\alpha,\beta\right)}\left(z\right)=K_N^{\left(\alpha ,\beta\right)}\left(z\right) \log\left(\frac{1-z}{1+z}\right) \\
+ 4 \sum _{i=0}^{\left[\frac{N-1}{2}\right]} \frac{K_{N-2i-1}^{(\alpha ,\beta )}(z)}{2
i+1}\; ,\; \;  \left(\alpha,\beta = \pm 1/2\right)
\end{gathered}
\label{w8}
\end{equation}
to the form for all cases, except in the case $\alpha=\beta=-1/2$ (see Eq.~(\ref{w7t}) in
Appendix \ref{pril-b}).

\subsection{Hypersingular variant}
\label{hyper}

Consider a hypersingular variant of integral  (\ref{eqint13}), where
$$
g(t,z)=1/(t-z)^{2}\; .
$$

The concept of calculation of the finite part of hypersingular integral was first put
forward by Hadamard (J. Hadamard, Lectures on Cauchy's Problem in Linear Partial
Differential Equations, Yale University Press, 1923) and developed in \cite[et
al.]{NME:NME499,Kaya1987,Kutt1975}. The finite part of hypersingular integral marked by
the sign $\intlh$ is related to the Cauchy integral  by the equation \cite{Kaya1987}
\begin{equation}
\inth \frac{w(t) F(t)}{\left(t-z\right)^{2}}\hskip 1pt \mathrm{d}t=
 \frac{\mathrm{d}}{\mathrm{d}z}\left[\intc\frac{w(t) F(t)}{t-z}\hskip 1pt
 \mathrm{d}t\right], \; -1 < z < 1 .
 \label{tx2-1}
\end{equation}

Useful in applications can be a subtraction, in which the hypersingular version of the
equation (\ref{eqint9a}) is expressed as
\begin{multline}
\inth \frac{F(t)\hskip 1pt w(t)}{\left(t-z\right)^{2}}\hskip 1pt \mathrm{d}t =\intc
\left(F(t)-F(z)\right) \frac{w(t)}{\left(t-z\right)^{2}}\hskip 1pt \mathrm{d}t\\+
F(z)\inth \frac{ w(t) }{\left(t-z\right)^{2}}\hskip 1pt \mathrm{d}t\; . \label{hsi4}
\end{multline}

Correspondingly, the weight factors of the quadrature formula
\begin{equation}
\inth\frac{w(t) F(t)}{\left(t-z\right)^{2}}\hskip 1pt \mathrm{d}t =
\\ \sum\limits_{i=1}^{N} \omega_{i}^{H}\left(z\right)F(\xi _{i,N}) \label{tx2-2}
\end{equation}
are related to coefficients (\ref{fz3}) by the relation
\begin{multline}
\omega_{i}^{H}\left(z\right)=
\frac{\mathrm{d}}{\mathrm{d}z}\left[\omega_{i}^{C}\left(z\right)\right]
\\
= \frac{1}{P_{N}^{\hskip 1pt \prime \left(\alpha,\beta\right)}\left(\xi
_{i,N}\right)}\frac{\mathrm{d}}{\mathrm{d}z}\left[\widetilde{\omega}_{N}^{C}\left(z,\xi
_{i,N}\right)\right]\; .\label{tx2-3}
\end{multline}

Then the weight factors of integral (\ref{tx2-2}) can be calculated by the formulas
\begin{widetext}
\begin{equation}
\omega_{i}^{H}\left(z\right) =\frac{1}{P_{N}^{\hskip 1pt \prime
\left(\alpha,\beta\right)}\left(\xi _{i,N}\right)} \left\{
\begin{array}{ll}
\dfrac{\Pi_{N}^{\hskip 1pt \prime
\left(\alpha,\beta\right)}\left(z\right)}{\left(z-\xi_{i,N}\right)}-\dfrac{\Pi_{N}^{\left(\alpha,\beta\right)}\left(z\right)-
\Pi_{N}^{\left(\alpha,\beta\right)}\left(\xi
_{i,N}\right)}{\left(z-\xi_{i,N}\right)^{2}}&\;, z\neq \xi _{i,N}\; ,
\hfill \\
\dfrac{1}{2}{\Pi_{N}^{\hskip 1pt \prime \prime \left({\alpha,\beta} \right)}\left(\xi
_{i,N}\right)}&\;,  z=\xi _{i,N}\; . \label{tx2-4}
\hfill \\
\end{array}
\right.
\end{equation}
\end{widetext}
It is worth noting that this formula for hypersingular integral has been obtained for the
first time.

\subsubsection{$w(t)=\sqrt{\left(1+t\right)/\left(1-t\right)}$}

For the Cauchy integral at $\alpha=-\beta=-1/2$, one obtains
\begin{eqnarray}
&&\Pi_{n}^{\left(-1/2,1/2\right)}\left(z\right) \nonumber \\
&&=\int_{-1}^1 \sqrt{\frac{1+t}{1-t}} \frac{V_n(t)}{(t-z)} \mathrm{d}t = \pi W_{n}(z) \;
, \label{vn-1}
\end{eqnarray}
where $V_{n}(z)$ and $W_{n}(z)$ are the Chebyshev polynomials of the third and fourth
order, correspondingly  \cite{Mason2002} (see Appendix \ref{pril-b}).

Then the quadrature formula for hypersingular integral has the form
\begin{equation}
\inth \sqrt{\frac{1+t}{1-t}} \hskip 1pt \frac{F(t)}{(t-z)^{2}}\mathrm{d}t \approx
\sum\limits_{i = 1}^N \omega_{i}^{HV}\left(z\right) F\left(\xi _{i,N}\right) ,
\label{vn-7}
\end{equation}
where
\begin{multline}
\omega_{i}^{HV}\left(z\right) =\frac{\pi}{V_N^{\prime}(\xi_{i,N})}
\\
\times\left\{
\begin{array}{cl}
\dfrac{
W_{N}^{\prime}\left(z\right)}{\left(z-\xi_{i,N}\right)}-\dfrac{W_{N}\left(z\right)-W_{N}\left(\xi
_{i,N}\right)}{\left(z-\xi_{i,N}\right)^{2}}&\;,z\neq \xi _{i,N}\; ,
\hfill \\
\dfrac{1}{2}{W_N^{\hskip 1pt \prime \prime}(\xi_{i,N})}&\;,z=\xi _{i,N}\; .
\end{array}
\right. \label{vn-8}
\end{multline}

\subsubsection{$w(t)=\sqrt{1-t^2}$}

The quadrature formula for a hypersingular integral with weight function
$w(t)=\sqrt{1-t^2}$ are readily determined in a similar way from the Eq.~ (\ref{tx2-4})

The weight factors  for  hypersingular integral with a weight function
$w(t)=\sqrt{1-t^2}$ are readily determined from Eq.~(\ref{tx2-4}) and relationship
\begin{multline}
\inth \cfrac{T_n(t)}{\sqrt{1-t^2}\hskip 1pt (t-z)^2}\hskip 2pt \mathrm{d}t \\  =
\left\{\begin{array}{cc}
0&\;, n=0,1 \; , \\
\pi U^{\prime}_{n-1}(z)=2 \pi \hskip 1pt C^{(2)}_{n-2}(z) &\;, n \geqslant 2 \; ,  \\
\end{array}\right. \label{p9tt}
\end{multline}
where $C^{(\alpha)}_{n}(z)$ are Gegenbauer polynomials.

The appropriate  quadrature, takes the form
\begin{equation}
\inth \frac{F(t)}{(t-z)^{2}\sqrt{1-t^2}}\mathrm{d}t \approx \sum\limits_{i = 1}^N
\omega_{i}^{HT}\left(z\right) F\left(\xi _{i,N}\right)\; , \label{tn3}
\end{equation}
where
\begin{multline}
\omega_{i}^{HT}\left(z\right) =\frac{\pi}{N U_{N-1}\left({\xi _{i,N}} \right)}
\\
\times\left\{
\begin{array}{cl}
\dfrac{2\hskip 1pt
C_{N-2}^{(2)}\left(z\right)}{\left(z-\xi_{i,N}\right)}-\dfrac{U_{N-1}\left(z\right)-U_{N-1}\left(\xi
_{i,N}\right)}{\left(z-\xi_{i,N}\right)^{2}}&\;,z\neq \xi _{i,N}\; ,
\hfill \\
4\hskip 1pt C_{N-3}^{(3)}\left(\xi_{i,N}\right)&\;,z=\xi _{i,N}\; .
\end{array}
\right. \label{tn4}
\end{multline}
or
\begin{equation}
\omega_{i}^{HT}\left(z\right) = \frac{4\pi}{N}\sum\limits_{k = 2}^{N-1}
T_k\left(\xi_{i,N}\right) C_{k-2}^{(2)}\left(z\right)\; ,\label{tn5}
\end{equation}
where
\begin{equation}
T_k\left(\xi_{i,N}\right) = \cos[\left(k-1/2\right)\pi/N]\; . \label{zTn}
\end{equation}

\subsubsection{Special case}
\label{logsing4}

In practice, a quadrature formula with subtraction can be useful
\begin{equation}
\intc  \frac{F(t)-F(z)}{\left(t-z\right)^{2}} \hskip 1pt \mathrm{d}t \approx
\sum\limits_{i=1}^{N} {\omega}_{i}^{HS}\left(z\right)F(\xi _{i,N})\; . \label{qm1}
\end{equation}

Using (\ref{eqint16}), we find a formula for calculating weight factor with the
singularity ${\omega}_{i}^{HS}\left(z\right)$  in the form
\begin{equation}
{\omega}_{i}^{HS}\left(z\right) = \lambda_{i,N}^{(\alpha,\beta)}\sum\limits_{m =0}^{N- 1}
{ \frac{1}{{h}_m} P_m^{\left( {\alpha ,\beta } \right)} \left( {\xi _{i,N} }
\right)}\hskip 1pt J^{H}_{m}(z)\; ,\label{qm2}
\end{equation}
where
\begin{equation}
J_{m}^{H}(z) =\int_{-1}^1
\frac{P_m^{\left(\alpha,\beta\right)}\left(t\right)-P_m^{\left(\alpha,\beta\right)}\left(z\right)}{\left(t-z\right)^{2}}\hskip
1pt \mathrm{d}t \; .\label{qm3}
\end{equation}
Using the identity
\begin{equation}
\frac{F(t)-F(z)}{\left(t-z\right)^{2}}=
\frac{\mathrm{d}}{\mathrm{d}z}\left[\frac{F(t)-F(z)}{t-z}\right]+\frac{F^{\prime}(z)}{t-z}
\label{qm4}
\end{equation}
The integral (\ref{qm3}) is transformed to the form
\begin{multline}
J_{m}^{H}(z)=\int_{-1}^1
\frac{P_m^{\left(\alpha,\beta\right)}\left(t\right)-P_m^{\left(\alpha,\beta\right)}\left(z\right)}{t-z}\hskip
1pt \mathrm{d}t\\ + P_m^{\prime\hskip 1pt \left(\alpha ,\beta\right)}\left(z\right)
\log\left(\frac{1-z}{1+z}\right)\; . \label{qm5}
\end{multline}

This integral can be calculated in principle for arbitrary values of $\alpha$ and
$\beta$. When solving physical problems, we can restrict ourselves to $\alpha,\beta=\pm
1/2$. Here we give the formula for the case $\alpha=\beta=-1/2$
\begin{equation}
\omega_i^{HS}(z) = \frac{2}{N}\hskip 3pt  \sideset{}{ \hskip 1pt ^{\prime}}\sum\limits_{m
= 1}^N{T_{m-1}}({\xi _{i,N}})\hskip 1pt J_{m-1}^{H}(z)\;, \label{qm6}
\end{equation}
where
\begin{multline}
 J_{m}^{H}(z)=m \hskip 1pt {U_{m - 1}}(z)\hskip 1pt \log\left(\frac{1 - z}{1 + z} \right) \\
+ 4 \sum _{j=0}^{b_m}\left(\frac{m}{2j+1}-1\right)U_{m-2j-2}(z)\hskip 1pt c_{j}^{m}(b_m)
\; ,\\ b_m=\left[\frac{m-1}{2}\right] \; .\label{qm7}
\end{multline}
the presence of a function $c_{j}^{m}(n)$ (\ref{af-2})  indicate that the last term in
the sum is divided by two, if $m$ is an odd number.

Eqs.~(\ref{tx2-4}), (\ref{vn-8}), (\ref{tn4}) and  (\ref{qm6}) for the weight factors
makes it possible to calculate them with high accuracy and, correspondingly, it can be
used for solving the Schr\"{o}dinger equation with linear potential in momentum space.

\subsection{Logarithmic singularity}
\label{logsing3}

Let us consider weight coefficient (\ref{eqint13}) for polynomials
${K}_{n}^{\left(\alpha,\beta\right)}\left(z\right)$ (\ref{w5dop}), when $g(t,z)\sim
\log\left|t-z\right|$ and  $w(t)=1,\sqrt{(1+t)/(1-t)}$.

\subsubsection{Variant $w(t)=1$}
\label{logsing1}

Let us consider the singular function  of the form
\begin{equation}
g(t,z)=\log\left|t-z\right| \; .\label{gtz1}
\end{equation}

Using (\ref{eqint13}),(\ref{eqint13z}) and (\ref{w5}), we find a formula for calculating
weight factor with the logarithmic singularity $\omega_{i}^{\mathrm{log}}(z)$   in the
form
\begin{equation}
\omega_{i}^{\mathrm{log}}\left(z\right)= \frac{2}{K_{N}^{\hskip 1pt
\prime\left({\alpha,\beta} \right)} \left(\xi _{i,N}\right)} \sum\limits_{m=0}^{N-1}
K^{\left({\alpha,\beta} \right)}_{N-1-m}(\xi _{i,N})\hskip 1pt J_{m}^{U}(z)\;
.\label{sp-3k}
\end{equation}
where
\begin{equation}
J_{m}^{U}(z)=\int_{-1}^1 \log\left|t-z\right| U_m\left(t\right) \hskip 1pt \mathrm{d}t
\label{sp-4k}\; .
\end{equation}
Thus, the analytical evaluation of the coefficients $\omega_{i}^{\mathrm{log}}(z)$
depends on the possibility of an analytic calculation of the integral (\ref{sp-4k}).

Using integration by parts of the integral (\ref{sp-4k}), we write in the form
\begin{multline}
J_{m}^{U}(z)= \frac{1}{m+1} \left.\bigg[ (-1)^m \log(z+1) + \log(1-z)\right.\\
\left.-\int_{-1}^1 \frac{T_{m+1}\left(t\right)}{t-z} \hskip 1pt \mathrm{d}t \right.\bigg]
\; . \label{sp-5k}
\end{multline}

The Cauchy integral in the equation (\ref{sp-5k}) can be calculated using (\ref{w8})
\begin{multline}
\intc\frac{T_{m+1}(t)}{(t-z)} \mathrm{d}t  = T_{m+1}\left(z\right)
\log\left(\frac{1-z}{1+z}\right) \\
 +4\sum\limits_{k=0}^{\left[\frac{m}{2}\right]}
\frac{T_{m-2 k}(z)}{2k+1}c_{k}^{m}\left([m/2]\right)  \;,  \label{sp-6k}
\end{multline}
where the function $c_{k}^{m}\left(n\right)$ is defined by Eq.~(\ref{af-2})

As a result, for the integral  with a logarithmic singularity (\ref{gtz1}), we obtain the
quadrature formula
\begin{equation}
\int_{-1}^{1} \log\left|t-z\right|\hskip 1pt F(t)\mathrm{d}t \approx \sum\limits_{i =
1}^N \omega_{i}^{\mathrm{log}}\left(z\right) F\left(\xi _{i,N}\right)\; ,
\label{vn-12tn1}
\end{equation}
where
\begin{multline}
\omega_{i}^{\mathrm{log}}\left(z\right)= \frac{2}{K_{N}^{\hskip 1pt
\prime\left({\alpha,\beta} \right)} \left(\xi _{i,N}\right)} \sum\limits_{m=0}^{N-1}
\frac{K^{\left({\alpha,\beta} \right)}_{N-1-m}(\xi _{i,N})}{m+1} \\
\times \left. \Bigg \{(-1)^m \log(z+1) + \log(1-z) \right. \\
- T_{m+1}\left(z\right) \log\left(\frac{1-z}{1+z}\right)
\\
\left. - 4\sum\limits_{k=0}^{\left[\frac{m}{2}\right]}\frac{T_{m-2 k}(z)}{2k+1}
c_{k}^{m}\left([m/2]\right) \right.\Bigg\}\; .
\label{sp-7k}
\end{multline}
If $\alpha=\beta=-1/ 2$, then the summation in (\ref{sp-7k}) (index $m$) the last term is
divided by two.

Structure-analogous coefficients were obtained for $T_N(t)$ polynomials in
\cite{Deloff:2006hd} and used to solve the Schr\"{o}dinger equation with the Coulomb
potential in momentum space \cite{Deloff:2006xx}.

\subsubsection{Variant $w(t)=w^{(\alpha,\beta)}(t)=\sqrt{(1+t)/(1-t)}$}
\label{logsing1v}

The Chebyshev polynomial $V_n(t)$ of the third kind is  defined by \cite{Mason2002}
\begin{equation}
V_n \left(t\right)=\frac{\cos\left[\left(n+1/2\right)\arccos(t)
\right]}{\cos\left[\arccos(t)/2\right]} \; .\label{f18}
\end{equation}
Hence, the zeros of $V_n(t)$ occur at (Table \ref{tabT-1} in Appendix \ref{pril-b})
\begin{multline}
\xi_{i,N} =\cos\theta_{i,N}\\
= \cos\left(\frac{2\hskip 1pt i -1}{2\hskip 1pt
N+1}\right)\;,\; \; \; \left( {i = 1,\ldots ,N} \right)\; .\label{fc19}
\end{multline}

It is easy to deduce from (\ref{eqint17}) that the  function
$\lambda_{i,N}^{\left(-1/2,1/2\right)}$ is
\begin{equation}
\lambda_{i,N}^{\left(-1/2,1/2\right)} =\frac{4\pi}{2
N+1}\cos^2\left[\frac{\theta_{i,N}}{2}\right]\; . \label{f7cv}
\end{equation}
After simple calculations from the general formula (\ref{eqint16}) we obtain an equation
for the weight factors in the case when $\alpha=-\beta=1/2$
\begin{multline}
\omega_{i}^{{V}}\left(z\right) = \frac{4 }{2\hskip 1pt N+1} \cos
\left(\frac{\theta_{i,N}}{2}\right) \sum\limits_{m = 0}^{N -
1}\cos\left[\left(m+\frac{1}{2}\right) \theta_{i,N}\right]
\\ \times \int_{ - 1}^1\sqrt{\frac{1+t}{1-t}} \hskip 1pt
\log\left|t-z\right| V_m \left(t\right)\mathrm{d}t\; . \label{f8cnv}
\end{multline}

Using
\begin{equation}
\sqrt{\frac{1+t}{1-t}}\hskip 1pt
V_{k}(t)=\frac{T_{k+1}\left(t\right)+T_{k}\left(t\right)}{\sqrt{1-t^2}} \label{vn-10}
\end{equation}
it  follows, from the series for the logarithmic function
\begin{multline}
 \log\left|t-z\right|
 = -\log 2 \\ -2 \hskip 1pt  \sum\limits_{m =
1}^{\infty}\frac{1}{m}\hskip 1pt T_m(z) \hskip 1pt T_m(t)\; , \hskip 1pt\; -1 \leqslant
z,t \leqslant 1 \label{vn-11}
\end{multline}
that
\begin{eqnarray}
&& \int_{-1}^1 \log\left|t-z\right| \hskip 1pt \hskip 1pt \sqrt{\frac{1+t}{1-t}}
V_k\left(t\right) \hskip 1pt
\mathrm{d}t \nonumber \\
&& = \left\{
\begin{array}{ll}
 -\pi \left(\cfrac{T_k(z)}{k}+\cfrac{T_{k+1}(z)}{k+1}\right) &\; ,  k \geqslant 1\; , \\
 -\pi\hskip 1pt\left(\log 2+ z\right)&
\; ,   k=0 \;.   \\
\end{array}\right.
\; \label{vn-12}
\end{eqnarray}

Then for integrals with a logarithmic singularity (\ref{gtz1}) we obtain the following
quadrature formula
\begin{equation}
\int_{-1}^{1} \log\left|t-z\right|\hskip 1pt \sqrt{\frac{1+t}{1-t}} \hskip 1pt
F(t)\mathrm{d}t \approx \sum\limits_{i = 1}^N \omega_{i}^{{V}}\left(z\right) F\left(\xi
_{i,N}\right)\hskip 1pt  , \label{vn-13}
\end{equation}
where
\begin{eqnarray}
&&\omega_{i}^{{V}}\left(z\right) = -\hskip 1pt \frac{4\pi}{2\hskip 1pt N+1} \cos
\left(\frac{\theta_{i,N}}{2}\right)\left.\Bigg[\left(\log 2 +z\right)\cos
\left(\frac{\theta_{i,N}}{2}\right)  \right. \nonumber \\
&& + \sum\limits_{m =1}^{N -1}\cos\left[\left(m+\frac{1}{2}\right) \theta_{i,N}\right]
\left. \left(\cfrac{T_k(z)}{k}+\cfrac{T_{k+1}(z)}{k+1}\right)\right]\hskip 1pt .
 \label{vn-15}
\end{eqnarray}

\subsubsection{Variant $w(t)=\sqrt{1-t^2}$}
\label{logsing1tn}

For convenience, we consider not only the case $\alpha=-\beta=1/2$  but also the case
when $\alpha=\beta=-1/2$. In this case the weight factors of the quadrature formula for
integrals with a logarithmic singularity of the form
\begin{equation}
\int_{-1}^{1} \log\left|t-z\right|\frac{F(t)}{\sqrt{1-t^2}}\mathrm{d}t \approx
\sum\limits_{i = 1}^N \omega_{i}^{T}\left(z\right) F\left(\xi _{i,N}\right)\; ,
\label{tn6}
\end{equation}
are obtained from the Eq.~(\ref{eqint16}) and can be written in the form
\begin{equation}
\omega_{i}^{T}\left(z\right) = - \frac{\pi}{N} \left[\ln 2 +2\hskip 1pt \sum\limits_{k =
1}^{N-1} \frac{1}{k} \; T_k\left(\xi_{i,N}\right) T_{k}\left(z\right)\right]\;
.\label{tn7}
\end{equation}

\subsubsection{Special case}
\label{logsing2}

Let us consider weight coefficient (\ref{eqint13}) when
\begin{multline}
g(t,z)=Q_{0}\left(t,z\right)\\
=\log\left|\frac{1- t
z+\sqrt{\left(1-t^2\right)\left(1-z^2\right)}}{t-z}\right|\label{sp-1}
\end{multline}
and $\alpha, \beta = \pm 1/2$ , $w(t)=1$, i.e.,
\begin{equation}
\omega_{i}^{Q_0}\left(z\right)= \frac{1}{P_{n }^{\hskip 1pt \prime\left({\alpha,\beta}
\right)} \left( {\xi _{i,N} } \right)} \int_{-1}^1 Q_{0}\left(t,z\right) \hskip 1pt
\hskip 1pt \frac{P_N^{\left(\alpha ,\beta\right)} \left(t\right)}{t - \xi _{i,N}}\hskip
1pt \mathrm{d}t\; .\label{sp-2}
\end{equation}

An analytical calculation of (\ref{sp-2}) can be performed using  the Eq.~ (\ref{w5})
with polynomials $K_{N}^{\left({\alpha,\beta}\right)}\left(z\right)$. Using (\ref{w5}),
expression (\ref{sp-2}) takes the form
\begin{multline}
\omega_{i}^{Q_0}\left(z\right)= \frac{2}{K_{N}^{\hskip 1pt \prime\left({\alpha,\beta}
\right)} \left( {\xi _{i,N} } \right)} \sum\limits_{k=0}^{N-1} K_{N-1-k}^{\left(\alpha
,\beta\right)}(\xi _{i,N})\\
\times  \int_{-1}^1 Q_{0}\left(t,z\right)U_k\left(t\right) \hskip 1pt \mathrm{d}t\;
.\label{sp-3}
\end{multline}

The integral in (\ref{sp-3}) is calculated after integration by parts
\begin{multline}
\int_{-1}^1 Q_{0}\left(t,z\right)U_k\left(t\right) \hskip 1pt \mathrm{d}t =
\frac{\sqrt{1-z^2}}{k+1}\int_{-1}^1 \frac{T_{k+1}\left(t\right)}{\sqrt{1-t^2}\hskip 1pt
\left(t-z\right)}\hskip 1pt
\mathrm{d}t\\
= \frac{\pi \sqrt{1-z^2}}{k+1} U_{k}\left(z\right) \; .\label{sp-4}
\end{multline}

As a result, the quadrature formula for the integral with logarithmic singularity of type
(\ref{sp-1})
\begin{multline}
\int_{-1}^{1}F(t)\hskip 1pt \log\left|\frac{1- t
z+\sqrt{\left(1-t^2\right)\left(1-z^2\right)}}{t-z}\right| \hskip 1pt \mathrm{d}t
\\ \approx \sum\limits_{i=1}^{N} \omega_{i}^{Q_0}\left(z\right) F(\xi _{i,N})
\label{logtx2-2}
\end{multline}
contains the weight factors
\begin{multline}
\omega_{i}^{Q_0}\left(z\right)=\\
\frac{2 \pi \; \sqrt{1-z^2}}{K_{N}^{\hskip 1pt \prime\left({\alpha,\beta} \right)} \left(
{\xi _{i,N} } \right)} \sum\limits_{k=0}^{N-1} K_{N-1-k}^{\left(\alpha ,\beta\right)}(\xi
_{i,N})\hskip 1pt \frac{U_{k}\left(z\right)}{k+1} \hskip 1pt .\label{sp-5}
\end{multline}

The  weight coefficients (\ref{sp-7k}), (\ref{vn-15}),(\ref{tn7})  and  (\ref{sp-5})
despite the cumbersome form, can be calculated with a sufficient degree of accuracy and
used to solve the equations.

\section{Energy spectrum for Coulomb potential}
\label{sect-5}

The equation with Coulomb potential
\begin{multline}
\frac{k^2}{2\mu}\phi_{n \ell}(k)-\frac{\alpha}{\pi\hskip 1pt k} \int_{0}^{\infty}
Q_{\ell}(y)\hskip 1pt k^{\prime}\hskip 1pt \phi_{n \ell}(k^{\prime})
\mathrm{d}k^{\prime}=E_{n \ell}\phi_{n \ell}(k) \label{eq1v}
\end{multline}
we transform to the form
\begin{equation}
\tilde{k}^2\hskip 1pt \phi_{n \ell}(\tilde{k})-\frac{2}{\pi\hskip 1pt \tilde{k}}
\int_{0}^{\infty} Q_{\ell}(y)\hskip 1pt \tilde{k}^{\prime} \phi_{n
\ell}(\tilde{k}^{\prime}) \mathrm{d}\tilde{k}^{\prime}=\varepsilon_{n \ell}\hskip 1pt
\phi_{n \ell}(\tilde{k})\hskip 1pt , \label{eq1vv}
\end{equation}
where
\begin{equation}
\begin{array}{cc}
k=\beta \tilde{k}\; , & \phi_{n \ell}(\tilde{k})= \beta^{3/2} \phi_{n \ell}({k})\; , \\
\beta=\mu \hskip 1pt \alpha  \; , & E_{n \ell}=\cfrac{\beta^{2}}{2 \mu}\hskip 2pt \varepsilon_{n \ell}\;. \\
\end{array}
\label{pr1h}
\end{equation}

In the case of Coulomb potential, the exact values of energies are known, namely,
\begin{equation}
\varepsilon_{n \ell}^{C}=-1/n^2 . \label{encoul}
\end{equation}

The accuracy of solving the equation  will be determined using the relative error
\begin{equation}
\delta_{n \ell}=\left|\frac{\varepsilon_{n \ell}-\varepsilon_{n
\ell}^{(N)}}{\varepsilon_{n \ell}}\right| \; , \label{deltotn}
\end{equation}
where $\varepsilon_{n \ell}$ are exact eigenvalues and $\varepsilon_{n \ell}^{(N)}$ is
the energy spectrum obtained by the numerical solution of the eigenvalues problem for
matrix $H$ at the given number of $N$
\begin{equation}
\sum\limits_{j=1}^{N}\hskip 1pt H_{i j}\phi_{n \ell}(\xi_{j,N})=\varepsilon^{(N)}_{n
\ell} \phi_{n \ell}(\xi_{i,N})\; .\label{eq5v}
\end{equation}

The calculations were carried out in the Wolfram Mathematica system \cite{Wolfram2017},
and the chosen accuracy of the weight factors and zeros was equal to $90$.  For all
calculations, we assume that numeric parameter $\beta_0=0.999992$.

\subsection{Quadrature rules  for $\ell \geqslant 0$}

For the numerical solution of the Schr\"{o}dinger equation (\ref{eq1v}) with the
logarithmic singularity, we use three realizations of the eigenvalue problem with the
help of  quadrature rules.

In the first method (Method I) we use the Chebyshev polynomials of the third kind
$V_n(t)$ with the function $w(t)=\sqrt{(1 + t)/(1-t)}$ and, respectively, the weight
factors (\ref{vn-15}) to eliminate the logarithmic singularity. The second method (Method
II) includes using Chebyshev polynomials of the first kind $T_n(t)$ with the function
$w(t) = 1$ and weight factors (\ref{sp-7k}) for integrals with a logarithmic singularity.
In Method III, we apply the  quadrature rule with weights factors (\ref{eqint49s}) to all
integrals  in the subtracted integral equation (\ref{prh3}) (Land\'{e} subtraction
method). Some characteristics of the methods  are presented in the Table \ref{tab-1n}.
\begin{table}[h p t b]
\caption{Characteristics of methods.}
\begin{ruledtabular}
\begin{tabular}{ccccc}
  Method  & $P_{n}^{\left({\alpha,\beta} \right)} \left(t\right)$ & $\xi_{i,n}$ &   $\omega_{i}$& \\
  \hline
  I     & $V_{n}(t)$ & $\cos\left(\cfrac{2\hskip 1pt i-1}{2n+1}\hskip 1pt \pi\right)$  & $\omega_{i}^{V}\left(z\right)$, & (\ref{vn-15})  \\
  II    & $T_{n}(t)$ & $\cos\left(\cfrac{i-1/2}{n}\pi\right)$                         & $\omega_{i}^{\mathrm{log}}\left(z\right)$,    &(\ref{sp-7k})                    \\
  III   & $T_{n}(t)$ & $\cos\left(\cfrac{i-1/2}{n}\pi\right)$                         & $\omega_{i}^{\mathrm{st}}$,              &(\ref{eqint49s})           \\
\end{tabular}
\end{ruledtabular}
\label{tab-1n}
\end{table}

By making use of mapping (\ref{an3})
\begin{equation}
\tilde{k}= \beta_{0}{\frac{1+z}{1-z}}\; ,\; \; \;  \tilde{k}^{\prime}=
\beta_{0}{\frac{1+t}{1-t}}\; ,\label{kt3w}
\end{equation}
we transform Eq.~ (\ref{eq1vv}) to
\begin{multline}
\frac{4 \beta_{0}}{\pi}\hskip 1pt {\frac{1-z}{1+z}} \int_{-1}^{1} Q_{\ell}(y(z,t))
\hskip 1pt \left(\frac{1+t}{1-t}\right) \phi_{n \ell}(t) \frac{\mathrm{d}t}{(1-t)^2}\\
=\left(\beta_{0}^{2}\left(\frac{1+z}{1-z}\right)^{2}- \varepsilon_{n \ell}\right) \hskip
1pt \phi_{n \ell}(z) \;, \label{eq3v}
\end{multline}
where
\begin{equation}
Q_{\ell}(y(z,t))=P_{\ell}(y(z,t))\log\left|\frac{1-t \hskip 1pt z}{t-z}\right| -
w_{\ell-1}(y(z,t))\; , \label{eq5zt}
\end{equation}
and
\begin{equation}
y(z,t)=\frac{2(t- z)^2}{\left(1-t^2\right) \left(1-z^2\right)}+1 \;.\label{pr2v}
\end{equation}
To shorten the notation in this section, we introduce functions
\begin{equation}
\bar{k}_{i} =\left(\frac{1+\xi_{i,N}}{1-\xi_{i,N}}\right)\,,\; \;  {\overline{dk}}_{i}
=\frac{1}{\left(1-\xi_{i,N}\right)^{2}}\,.\label{ki}
\end{equation}

Consider the numerical solution to Eq.~ (\ref{eq3v})  by means of the quadrature
formulas. Employing the  Method I and putting that $z=\xi_{i,N}$ and $t=\xi_{j,N}$, the
integral Eq.~ (\ref{eq3v}) can be approximated by the matrix equation (\ref{eq5v}) with
\begin{equation}
H_{i j}=\beta_{0}  \left[\beta_{0}\hskip 1pt \delta_{i, j}\hskip 1pt \bar{k}_{j}^{2}-
\dfrac{4}{\pi} \left(1/\bar{k}_{i}\right)\sqrt{\bar{k}_{j}}  Q_{\ell}^{V}(y_{i
j}){\overline{dk}}_{j} \right]\hskip 1pt ,\label{eq6vv}
\end{equation}
where
\begin{eqnarray}
&&Q_{\ell}^{V}(y_{i j})=\lambda_{j,N}^{\left(-1/2,1/2\right)} \nonumber \\
&\times& \left.\bigg[P_{\ell}(y_{i j})\log\left|1-\xi_{i,N} \hskip 1pt \xi_{j,N}\right| -
w_{\ell-1}(y_{i j})\right.\bigg]\nonumber \\
&-&\omega_{j}^{\mathrm{log}} \left(\xi_{i,N}\right)P_{\ell}(y_{i j})\; ,\; \;\; y_{i
j}=y\left(\xi_{i,N},\xi_{j,N}\right)\; .\label{eq7vb}
\end{eqnarray}
The weight factors $\lambda_{j,N}^{\left(-1/2,1/2\right)}$ and $\omega_{j}^{V}
\left(\xi_{i,N}\right)$ are determined by the Eqs.~ (\ref{f7cv}) and (\ref{vn-15}),
respectively, and the values of $\xi_{i,N} $ by the formula  (see Table \ref{tab-1n})
\begin{equation}
\xi_{i,N}= \cos\left(\dfrac{2\hskip 1pt i-1}{2 N + 1}\hskip 1pt \pi\right)\; .\label{vnw}
\end{equation}

Calculations using the  Method $\mathrm{II}$  adduce to a matrix of the form
\begin{equation}
H_{i j}=\beta_{0}  \left[\beta_{0}\hskip 1pt\delta_{i, j}\hskip 1pt \bar{k}_{j}^{2}-
\dfrac{4}{\pi} \left(\bar{k}_{j}/\bar{k}_{i}\right) \hskip 1pt Q_{\ell}^{T}(y_{i
j})\hskip 1pt {\overline{dk}}_{j}\right]\,,\label{eq7vvz}
\end{equation}
where
\begin{eqnarray}
Q_{\ell}^{T}(y_{i j})&=&\omega_{j}^{\mathrm{st}}  \left.\bigg[P_{\ell}(y_{i
j})\log\left|1-\xi_{i,N} \hskip 1pt \xi_{j,N}\right| -
w_{\ell-1}(y_{i j})\right.\bigg]\nonumber \\
&-&\omega_{j}^{\mathrm{log}} \left(\xi_{i,N}\right)P_{\ell}(y_{i j})\; .\label{eq7vc}
\end{eqnarray}
The weight factor $\omega_{j}^{\mathrm{st}}$ and $\omega_{j}^{\mathrm{log}}
\left(\xi_{i,N}\right)$ are determined by the Eqs.~ (\ref{eqint49s}) and (\ref{sp-7k}),
respectively, and the values of $\xi_{i,N} $ by the relationship
\begin{equation}
\xi_{i,N}=\cos \left(\dfrac{i-1/2}{N}\hskip 1pt \pi\right)\; . \label{tnzt}
\end{equation}

The matrix elements $H_{ij}$ of  Land\'{e}-subtracted integral equation (\ref{prh3})
with a Coulomb potential are
\begin{eqnarray}
H_{ii} &=& \beta_{0}  \left.\Bigg[\beta_{0}\hskip 2pt \bar{k}_{i}^{2} -\frac{2}{{\pi
}}\hskip 1pt
\mathcal{C}_{\ell}\hskip 1pt \bar{k}_{i} \right.\nonumber\\
&& \left. +\frac{4}{\pi}\sum\limits_{r=1}^{N} \omega_{r}^{\mathrm{st}} Q_{\ell}(y_{r i}
\ne 1) \left(\bar{k}_{i}/\bar{k}_{r}\right) \hskip 1pt {\overline{dk}}_{r}\right.\Bigg]
\,,
\nonumber\\
H_{ij} &=& - \frac{4 \beta_{0}}{\pi}\hskip 1pt \omega_{j}^{\mathrm{st}} \hskip 1pt\left(
\bar{k}_{j}/\bar{k}_{i}\right) \hskip 1pt
 Q_{\ell}(y_{i j}) \hskip 1pt
{\overline{dk}}_{j} \;  , \; \;  (i\ne j)\; . \label{eq8vc}
\end{eqnarray}
The diagonal matrix elements $H_{ii}$ of Eqs.~(\ref{eq6vv}), (\ref{eq7vvz}) are  finite
and all singularities are under control.

Numerical results calculated by three methods are compared with each other (see Table
\ref{tab:coulomb-1}).

\begin{table}[tb]
\centering \caption{Relative errors $\delta_{n \ell}$ (\ref{deltotn}) on the computed
Coulomb binding energies. Index $\mathrm{I,II,III}$ denotes that Methods
$\mathrm{I,II,III}$  are used for calculation of $\varepsilon_{n \ell}^{N}$ respectively
and  $7.1(-16)\equiv 7.1 \times 10^{-16}$.}
\begin{ruledtabular}
\begin{tabular}{llllll}
   &            &            &   $\ell=0$         &            &          \\
\hline
 {$N$} & \multicolumn{1}{c}{$n=1$} &
   \multicolumn{1}{c}{$n=2$} & \multicolumn{1}{c}{$n=3$} & \multicolumn{1}{c}{$n=4$} & \multicolumn{1}{c}{$n=5$} \\
\hline
 50$^{\mathrm{I}}$    & $2.3(-12)$  &  $2.3(-9)$   & $1.8(-8)$   & $7.5(-8)$   & $2.3(-7)$   \\
 100$^{\mathrm{I}}$   & $3.7(-14)$  &  $3.7(-11)$  & $2.9(-10)$  & $1.2(-9)$   & $3.7(-9)$   \\
 150$^{\mathrm{I}}$   & $3.3(-15)$  &  $3.3(-12)$  & $2.5(-11)$  & $1.1(-10)$  & $3.3(-10)$  \\
 150$^{\mathrm{II}}$  & $1.1(-16)$  &  $6.7(-15)$  & $1.1(-13)$  & $8.6(-13)$  & $4.1(-12)$  \\
 150$^{\mathrm{III}}$ & $7.1(-7)$   &  $1.4(-5)$   & $1.2(-4)$   & $5.3(-4)$   & $1.8(-3)$   \\
\hline
   &            &            &   $\ell=1$         &            &          \\
\hline
 {$N$} & \multicolumn{1}{c}{$n=1$} &
   \multicolumn{1}{c}{$n=2$} & \multicolumn{1}{c}{$n=3$} & \multicolumn{1}{c}{$n=4$} & \multicolumn{1}{c}{$n=5$} \\
\hline
 50$^{\mathrm{I}}$    & $2.2(-14)$  & $1.0(-12)$  & $1.4(-11)$  & $1.1(-10)$  & $4.4(-10)$  \\
 100$^{\mathrm{I}}$   & $2.3(-17)$  & $1.0(-15)$  & $1.5(-14)$  & $1.1(-13)$  & $5.8(-13)$  \\
 150$^{\mathrm{I}}$   & $4.0(-19)$  & $1.8(-17)$  & $2.6(-16)$  & $2.0(-15)$  & $1.0(-14)$  \\
 150$^{\mathrm{II}}$  & $4.7(-16)$  & $9.2(-15)$  & $8.7(-14)$  & $4.5(-13)$  & $1.8(-12)$  \\
 150$^{\mathrm{III}}$ & $4.2(-5)$   & $1.7(-4)$   & $3.6(-4) $  & $4.3(-4)$   & $1.2(-4)$   \\
 \hline
   &            &            &   $\ell=2$         &            &          \\
\hline
 {$N$} & \multicolumn{1}{c}{$n=1$} &
   \multicolumn{1}{c}{$n=2$} & \multicolumn{1}{c}{$n=3$} & \multicolumn{1}{c}{$n=4$} & \multicolumn{1}{c}{$n=5$} \\ \hline
 50$^{\mathrm{I}}$    & $2.2(-14)$  & $1.0(-12)$  & $1.4(-11)$  & $1.1(-10)$  & $4.4(-10)$  \\
 100$^{\mathrm{I}}$   & $2.3(-17)$  & $1.0(-15)$  & $1.5(-14)$  & $1.1(-13)$  & $5.8(-13)$  \\
 150$^{\mathrm{I}}$   & $4.0(-19)$  & $1.8(-17)$  & $2.6(-16)$  & $2.0(-15)$  & $1.0(-14)$  \\
 150$^{\mathrm{II}}$  & $1.7(-19)$  & $4.7(-18)$  & $6.1(-17)$  & $4.7(-16)$  & $2.6(-15)$ \\
 150$^{\mathrm{III}}$ & $1.4(-5)$   & $1.1(-4)$   & $4.8(-4)$   & $1.6(-3)$   & $4.3(-3)$  \\
\end{tabular}
\label{tab:coulomb-1}
\end{ruledtabular}
\end{table}
As follows from the results of the calculation, Methods $\mathrm{I}$ and $\mathrm{II}$
have excellent convergence with increasing $N$ and significantly exceed the accuracy of
Method $\mathrm{III}$. In addition, the accuracy of Methods $\mathrm{I}$ and
$\mathrm{II}$ increases with the increase of orbital number  $\ell$, unlike Method
$\mathrm{III}$.

Therefore, quadrature formulas (\ref{vn-15}) and (\ref{sp-7k}), in which the logarithmic
singularities of integrals are included into the weight factors, make it possible to
solve the Schr\"{o}dinger equation with Coulomb potential  in momentum space with high
accuracy.

\subsection{Special case for $\ell=0$}

Using mapping (\ref{an3x})
\begin{equation}
\tilde{k}= \beta_{0}\sqrt{\frac{1+z}{1-z}}\; ,\; \; \;  \tilde{k}^{\prime}=
\beta_{0}\sqrt{\frac{1+t}{1-t}}\; ,\label{kt3}
\end{equation}
Eq.~(\ref{eq1vv}) for $\ell=0$  is given by
\begin{multline}
\frac{2 \beta_{0}}{\pi}\hskip 1pt \sqrt{\frac{1-z}{1+z}}\\
\times\int_{-1}^{1} \phi_{n 0}(t)\hskip 1pt \log\left|\frac{1- t z+
\sqrt{\left(1-t^2\right)\left(1-z^2\right)}}{t-z}\right |\hskip 1pt \frac{\mathrm{d}t}{(1-t)^2} \\
=\left(\beta_{0}^{2}\left(\frac{1+z}{1-z}\right)-\varepsilon_{n 0}\right)\hskip 1pt
\phi_{n 0}(z) \;. \label{eq4v}
\end{multline}

Consider the numerical solution to Eq.~(\ref{eq4v})  by means of the quadrature formula
(\ref{logtx2-2}) with the weights (\ref{sp-5}). Using (\ref{logtx2-2}), integral equation
(\ref{eq3v}) reduces to the eigenvalues problem
\begin{equation}
\sum\limits_{j=1}^{N}\hskip 1pt H_{i j}\phi_{n 0}(\xi_{j,N})=\varepsilon^{(N)}_{n0}
\phi_{n 0}(\xi_{i,N})\; ,\label{eq5vvv}
\end{equation}
where the matrix elements of $H$ are given by the formula
\begin{multline}
H_{i j}=\beta_{0}  \left.\Bigg[\beta_{0}\hskip 2pt\delta_{i, j}\hskip 1pt \bar{k}_{j} -
\dfrac{2}{\pi} \dfrac{{\overline{dk}}_{j}}{\bar{k}_{i}^{1/2}}\hskip 1pt \omega_{j}^{Q_0}
\left(\xi_{i,N}\right) \right.\Bigg]\; .\label{eq6v}
\end{multline}
The matrix $\omega_{j}^{Q_0} \left(\xi_{i,N}\right)$ is calculated by means of
(\ref{sp-5}) and the functions  $\bar{k}_{i}$, $\overline{dk}_{i}$ are determined by the
Eq.~ (\ref{ki}).

We carry out the calculations for two sets of polynomials: the first-order Chebyshev
polynomials $T_n(t)$ ($\alpha=\beta=-1/2$) and the third-order Chebyshev polynomials
$V_n(t)$ ($\alpha=-1/2,~\beta=1/2$). The values of relative error (\ref{deltotn}),
obtained as a result of numerical solution, are given in Table \ref{tab-1a}, depending on
the number of nodes  $N$.
\begin{table}[h p t b]
\caption{The five relative errors $\delta_{n0}$  with $\ell = 0$ for polynomials
$V_n(t)$, obtained by solving Eq.~ (\ref{eq4v}).}
\begin{ruledtabular}
\begin{tabular}{cccccc}
  $N$  & $n=1$ & $n=2$ & $n=3$ & $n=4$ & $n=5$ \\
  \hline
  50    & $0.0(-90)$  & $9.5(-41)$  & $3.0(-21)$  & $4.6(-12)$  & $8.8(-7)$\\
  100   & $0.0(-90)$  & $3.0(-87)$  & $1.7(-49)$  & $1.1(-31)$  & $4.7(-21)$\\
  150   & $0.0(-90)$  & $0.0(-90)$  & $1.6(-78)$  & $2.0(-52)$  & $8.4(-37)$\\
\end{tabular}
\end{ruledtabular}
\label{tab-1a}
\end{table}

The solution of Eq.~ (\ref{eq4v}) for the first-order Chebyshev polynomials $T_n(t)$
leads to analogous results (see Table \ref{tab-2a}).
\begin{table}[h t p b]
\caption{Relative error $\delta_{n0}$ for polynomials $T_n(t)$.}
\begin{ruledtabular}
\begin{tabular}{cccccc}
  $N$  & $n=1$ & $n=2$ & $n=3$ & $n=4$ & $n=5$ \\
  \hline
  50   & $0.0(-90)$  & $8.0(-40)$   & $1.1(-20)$  & $1.1(-11)$  & $1.6(-6)$\\
  100  & $0.0(-90)$  & $2.7(-86)$   & $6.5(-49)$ & $2.8(-31)$   & $9.6(-21)$\\
  150  & $0.0(-90)$  & $ 0.0(-90)$  & $6.3(-78)$  & $5.3(-52)$  & $1.8(-36)$\\
\end{tabular}
\end{ruledtabular}
\label{tab-2a}
\end{table}

As follows from the results, ``nearly exact'' quadrature formula for the integral in the
Schr\"{o}dinger equation allows one to reproduce the energy spectrum $\varepsilon_{n 0}$
with a high degree of accuracy, greatly surpassing the analogous calculations
\cite{Deloff:2006xx,Tang:2001ii,Chen2013}.

\section{Results for the linear  potential}
\label{sect-6}

We write the Schr\"{o}dinger equation with linear confinement potential
\begin{equation}
\frac{k^2}{2\mu}\phi_{n\ell}(k)+\frac{\sigma}{\pi\hskip 1pt k^{2}} \int_{0}^{\infty}
Q_{\ell}^{\prime}(y)\hskip 1pt \phi_{n\ell}(k^{\prime})
\mathrm{d}k^{\prime}=E_{n\ell}\phi_{n\ell}(k) \;,
 \label{eq1d}
\end{equation}
in the form
\begin{equation}
\tilde{k}^2\hskip 1pt \phi_{n \ell}(\tilde{k})+\frac{1}{\pi \hskip 1pt \tilde{k}^{2}}
\int_{0}^{\infty} Q_{\ell}^{\prime}(y) \phi_{n \ell}(\tilde{k}^{\prime})
\mathrm{d}\tilde{k}^{\prime}=\varepsilon_{n \ell}\hskip 1pt \phi_{n \ell}(\tilde{k}) \;
\label{eq1dd}
\end{equation}
using the replacements
\begin{equation}
\begin{array}{lll}
  k=\beta \tilde{k}&\; , &  E=\cfrac{\beta^{2}}{2 \mu}\hskip 2pt \varepsilon\; , \\
  \beta=\left(2 \mu \hskip 1pt \sigma\right)^{1/3}&\; , & \phi_{n \ell}(\tilde{k})=
\beta^{3/2} \phi_{n \ell}({k}) \; .\\
\end{array}
\label{pr1d}
\end{equation}

We may  deduce from Eq.~(\ref{eq7}) and Land\'{e} subtraction term (\ref{contr1}), that
the Schr\"{o}dinger equation (\ref{eq1dd}) is
\begin{multline}
\left(\varepsilon_{n \ell} - \tilde{k}^2 \right) \phi_{n \ell}(\tilde{k})  \\
= \frac{1}{\pi \tilde{k}^2} \int_0^\infty
\left.\Big[\,Q^{\prime}_0(y)\left.\Big\{P_\ell(y)\phi_{n \ell}(\tilde{k}^{\prime}) -
\phi_{n\ell}(\tilde{k}) \right.\Big\} \right.\\ -\left. w^{\prime}_{\ell- 1}(y) \phi_{n\ell}(\tilde{k}^{\prime})\right.\Big] \mathrm{d} \tilde{k}^{\prime} \\
+ \frac{1}{\pi {\tilde{k}^2}} \int_0^\infty \,Q_0(y) P^{\prime}_\ell(y) \phi_{n
\ell}(\tilde{k}^{\prime}) \mathrm{d} \tilde{k}^{\prime} \; . \label{linp1}
\end{multline}

To test the accuracy of calculations of the energy spectrum, we use the equation
(\ref{deltotn}). In the particular case $\ell=0$ the exact result is known and the
binding energy is
\begin{equation}
\varepsilon_{n 0}^{L} =-z_{n}\; ,\; \; n=1, 2, 3 \ldots\; , \label{airy}
\end{equation}
where $z_{n}$ are zeroes of the Airy functions $\mathrm{Ai}(z)$.

In contrast to Coulomb potential, there are no exact analytical solutions with linear
potential for $\ell \geqslant 1$. For $\ell \geqslant 1$ the values marked as exact have
been computed by solving the  Schr\"{o}dinger equation (\ref{wfeq2}) in configuration
space. For this purpose we used the variational method of solving with trial
pseudo-Coulomb wave functions \cite{Fulcher:1993sk}
\begin{align}
\psi^{\mathrm{C}}_{n\ell} (\mathrm{r},\beta) &=
N_{n\ell}^{\mathrm{C}}\left(2\beta\right)^{3/2}\; (2\beta r)^{\ell}e^{ - \beta
\mathrm{r}}L_{n}^{2\ell + 2}(2\beta \mathrm{r})\;,\label{a4s} \\
 N_{n\ell}^{\mathrm{C}} &= \sqrt{{\frac{{n!}}{{(n + 2\ell + 2)!}}}}\;,\label{a4}
\end{align}
where $L_{n}^{\ell}(z) $ are the Laguerre polynomials with $n,\ell\geq 0$. In
\cite{Fulcher:1993sk}, the analytical expressions for the integrals with functions
(\ref{a4}) arising in coordinate space were obtained (see Appendix \ref{pril-3}). This
makes it possible to carry out calculations with a high degree of accuracy.

Therefore, the numerical solution in momentum space for $\ell \geqslant 1$ will be
compared to the solution of this equation in coordinate space.

\subsection{Quadrature rules $\ell \geqslant 0$}

To solve the Schr\"{o}dinger equation in a momentum space with a linear potential, we use
quadrature formulas (\ref{tn3}) and (\ref{qm1}). The methods of solving the equation with
the help of formulas (\ref{qm1}) and (\ref{tn3}) will be called as Method $\mathrm{A}$
and $\mathrm{B}$, respectively.

Let us now explain a method  of solution  (Method $\mathrm{A}$) of the integral equation
(\ref{linp1}). Employing the variable transformation (\ref{an3})
\begin{equation}
\tilde{k}= \beta_{0}{\left(\frac{1+z}{1-z}\right)}\; ,\; \; \;  \tilde{k}^{\prime}=
\beta_{0}{\left(\frac{1+t}{1-t}\right)}\; ,\label{kt3l}
\end{equation}
and then using quadrature relationships (\ref{tn3}) and  (\ref{qm1}) with the weight
factors (\ref{vn-12tn1}) and (\ref{qm6}), respectively, the subtracted integral equation
(\ref{linp1}) is approximated by the matrix equation (\ref{eq5v}), where the matrix
elements are
\begin{equation}
H_{i j}= \beta_{0}^{2} T_{i j}+\frac{1}{\beta_{0}\hskip 1pt \pi}
\left(1/\bar{k}^2_{i}\right)\left(V_{i j}^{\mathrm{H}}+V_{i j}^{\mathrm{Log}}\right)\; .
\label{meta}
\end{equation}
In Eq (\ref{meta})
\begin{eqnarray}
T_{i j}&=& \delta_{i,j}\hskip 1pt \bar{k}_{j}^{2}\; ,\nonumber \\
V_{i j}^{\mathrm{H}}&=& 2 \left.\bigg[\omega_j^{HS}(\xi_{i,N})P_{\ell}(y_{i j})Z_{i j}\right. \nonumber \\
&-& \delta_{i, j}\sum\limits_{k=1}^{N}\omega_k^{HS}(\xi_{j,N})Z_{k j}
\nonumber \\
&-&\left.\omega_{j}^{\mathrm{st}} \hskip
1pt w^{\prime}_{\ell- 1}(y_{i j})\right.\bigg]\overline{dk}_{j}\; ,\nonumber \\
V_{i j}^{\mathrm{Log}}&=& 2 \hskip 1pt P_{\ell}^{\prime}(y_{i
j})\left.\bigg[\omega_{j}^{\mathrm{st}}\log\left|1-\xi_{i,N} \hskip 1pt \xi_{j,N}\right|\right. \nonumber \\
&-&\omega_{j}^{\mathrm{log}} \left(\xi_{i,N}\right)\bigg]\overline{dk}_{j} \;
,\label{eqhij}
\end{eqnarray}
where
\begin{equation}
Z_{i j}= -\frac{1}{4}\left[\frac{\left(1-\xi _{i,N}^2\right)\left(1-\xi _{j,N}^2\right)}{
\left(1-\xi _{i,N} \xi _{j,N}\right)}\right]^{2} \; ,\label{zij}
\end{equation}
\begin{equation}
y_{ij}=\frac{2(\xi _{i,N}- \xi _{j,N})^2}{\left(1-\xi _{i,N}^2\right) \left(1-\xi
_{j,N}^2\right)}+ \delta_{i,j} \;.\label{yij}
\end{equation}
Weight factors $\omega_{j}^{\mathrm{st}}$,  $\omega_j^{HS}(\xi_{i,N})$ and
$\omega_{j}^{\mathrm{log}} \left(\xi_{i,N}\right)$ are determined by the Eqs.~
(\ref{eqint49s}), (\ref{qm6}) and (\ref{vn-12tn1}), respectively. The numbers $\xi_{i,N}$
are the zeros of the Chebyshev polynomial of the first kind $T_n (t)$ (see the
Eq.~(\ref{tnzt})).

Next we describe a quick  method of solution (Method $\mathrm{B}$)  of the hypersingular
integral equation (\ref{eq1dd}). The characteristic (specific) features of Method
$\mathrm{B}$ consist in using the change of variables (\ref{kt3}) and quadrature formulas
(\ref{tn3}) and (\ref{tn6}) with the weight function $w(t)=\sqrt{1-t^2} $ of the
Chebyshev polynomial $T_n(t)$.

As a result, the matrix $\widetilde {H}$ for calculating the energy spectrum using the
Method $\mathrm{B}$ is determined by the following relation
\begin{equation}
\widetilde{H}_{i j}= \beta_{0}^{2} \widetilde{T}_{i j}+\frac{1}{\beta_{0}\hskip 1pt \pi}
\left(1/\bar{k}_{i}\right)\left(\widetilde{V}_{i j}^{\mathrm{H}}+\widetilde{V}_{i
j}^{\mathrm{Log}}\right)\; . \label{metat}
\end{equation}
In Eq.~ (\ref{metat})
\begin{eqnarray}
\widetilde{T}_{i j}&=& \delta_{i, j}\hskip 1pt \bar{k}_{j}\; ,\nonumber \\
\widetilde{V}_{i j}^{\mathrm{H}}&=& \omega_j^{HT}(\xi_{i,N})P_{\ell}(y^{T}_{i
j})\left(\xi_{i,N}^2-1\right) \left(1+\xi_{j,N}\right)
\nonumber \\
&-&\frac{\pi}{N} \hskip
1pt w^{\prime}_{\ell- 1}(y^{T}_{i j})\; , \\
\widetilde{V}_{i j}^{\mathrm{Log}}&=&\left.\bigg[\frac{\pi}{N}\log\left|1-\xi_{i,N} \hskip 1pt \xi_{j,N}+\sqrt{1-\xi_{i,N}^2} \sqrt{1-\xi_{j,N}^2}\right|\right. \nonumber \\
&-&\omega_{j}^{{T}} \left(\xi_{i,N}\right)\bigg]\frac{ P_{\ell}^{\prime}(y^{T}_{i
j})}{\left(1-\xi_{j,N}\right)} \; ,\label{eqhtij}
\end{eqnarray}
where
\begin{equation}
y_{ij}^{T}=\frac{1-\xi_{i,N} \xi_{j,N}}{\sqrt{1-\xi_{i,N}^2} \sqrt{1-\xi_{j,N}^2}}
\;.\label{ytij}
\end{equation}
Weight factors  $\omega_j^{HT}(\xi_{i,N})$ and $\omega_{j}^{T} \left(\xi_{i,N}\right)$
are determined by the Eqs.~ (\ref{tn5}) and (\ref{tn7}), respectively.

The numerical results calculated by the Methods $\mathrm{A}$  and $\mathrm{B}$  are
compared with each other (see Table \ref{tab:linear-1}).
\begin{table}[tb]
\caption{Relative errors $\delta_{n \ell}$ (\ref{deltotn}) on the computed linear binding
energies. Index $\mathrm{A,B}$ denotes that Methods $\mathrm{A,B}$ are used for
calculation of $\varepsilon_{n \ell}^{N}$ respectively and  $7.6(-13)\equiv 7.6 \times
10^{-13}$.}
\begin{ruledtabular}
\begin{tabular}{llllll}
   &            &            &   $\ell=0$         &            &          \\
\hline
 {$N$} & \multicolumn{1}{c}{$n=1$} &
   \multicolumn{1}{c}{$n=2$} & \multicolumn{1}{c}{$n=3$} & \multicolumn{1}{c}{$n=4$} & \multicolumn{1}{c}{$n=5$} \\
\hline
 100$^{\mathrm{A}}$   & $ 2.9(-15)$   & $ 1.1(-14)$    & $ 2.2(-14)$    & $ 3.7(-14)$    & $ 5.4(-14)$   \\
 150$^{\mathrm{A}}$   & $ 1.1(-16)$   & $ 4.1(-16)$    & $ 8.6(-16)$    & $ 1.4(-15)$    & $2.1(-15)$\\
 150$^{\mathrm{B}}$   & $2.6(-27)$    & $ 6.2(-26)$    & $ 5.5(-24)$    & $ 8.8(-23)$    & $ 6.2(-22)$ \\
\hline
   &            &            &   $\ell=1$         &            &          \\
\hline
 {$N$} & \multicolumn{1}{c}{$n=1$} &
   \multicolumn{1}{c}{$n=2$} & \multicolumn{1}{c}{$n=3$} & \multicolumn{1}{c}{$n=4$} & \multicolumn{1}{c}{$n=5$} \\
\hline
 100$^{\mathrm{A}}$   & $8.0(-14)$   & $ 8.3(-14)$   & $ 1.6(-13)$   & $ 1.7(-13)$   & $ 2.4(-13)$  \\
 150$^{\mathrm{A}}$   & $7.0(-15)$   & $ 7.1(-15)$   & $ 1.4(-14)$   & $ 1.4(-14)$   & $ 2.1(-14)$  \\
 150$^{\mathrm{B}}$   & $2.6(-10)$   & $ 7.4(-10)$   & $ 1.4(-9)$    & $ 2.2(-9)$    & $ 3.1(-9)$  \\
 \hline
   &            &            &   $\ell=2$         &            &          \\
\hline
 {$N$} & \multicolumn{1}{c}{$n=1$} &
   \multicolumn{1}{c}{$n=2$} & \multicolumn{1}{c}{$n=3$} & \multicolumn{1}{c}{$n=4$} & \multicolumn{1}{c}{$n=5$} \\ \hline
 100$^{\mathrm{A}}$   & $6.5(-13)$  & $ 2.3(-13)$  & $ 5.6(-13)$  & $ 1.5(-12)$  & $ 8.0(-12)$  \\
 150$^{\mathrm{A}}$   & $5.5(-14)$  & $ 1.8(-14)$  & $ 9.9(-14)$  & $ 3.3(-14)$  & $ 1.4(-13)$  \\
 150$^{\mathrm{B}}$   & $1.5(-14)$  & $ 6.5(-14)$  & $ 1.7(-13)$  & $ 3.5(-13)$  & $ 6.2(-13)$ \\
 \hline
   &            &            &   $\ell=3$         &            &          \\
\hline
 {$N$} & \multicolumn{1}{c}{$n=1$} &
   \multicolumn{1}{c}{$n=2$} & \multicolumn{1}{c}{$n=3$} & \multicolumn{1}{c}{$n=4$} & \multicolumn{1}{c}{$n=5$} \\ \hline
 100$^{\mathrm{A}}$   & $3.2(-16)$   & $ 5.7(-15)$   & $ 1.1(-13)$  & $ 7.6(-13)$  & $ 1.8(-12)$  \\
 150$^{\mathrm{A}}$   & $3.8(-19)$   & $ 5.5(-19)$   & $ 1.2(-18)$  & $ 1.5(-18)$  & $ 3.3(-18)$  \\
 150$^{\mathrm{B}}$   & $6.1(-19)$   & $ 3.6(-18)$   & $ 1.2(-17)$   & $ 3.2(-17)$   & $ 6.9(-17)$\\
 \hline
   &            &            &   $\ell=4$         &            &          \\
\hline
 {$N$} & \multicolumn{1}{c}{$n=1$} &
   \multicolumn{1}{c}{$n=2$} & \multicolumn{1}{c}{$n=3$} & \multicolumn{1}{c}{$n=4$} & \multicolumn{1}{c}{$n=5$} \\ \hline
 100$^{\mathrm{A}}$   & $3.3(-18)$  & $ 9.9(-17)$  & $7.9(-17)$  & $ 1.0(-14)$ & $2.1(-13)$ \\
 150$^{\mathrm{A}}$   & $2.9(-24)$  & $ 4.1(-24)$  & $2.7(-22)$  & $ 3.2(-21)$ & $8.5(-20)$ \\
 150$^{\mathrm{B}}$   & $7.3(-20)$  & $5.4(-19)$   & $2.2(-18)$  & $6.8(-18)$  & $1.7(-17)$ \\
\end{tabular}
\end{ruledtabular}
\label{tab:linear-1}
\end{table}

As follows from the results of calculations, Method $\mathrm{A}$ gives a more accurate
result than method $\mathrm{B}$ for $\ell \geqslant 1$.

Thus, a special quadrature formula (\ref{qm1}) based on the use of a counter term and an
analytical calculation of weight factors involving a singularity gives a highly accurate
solution of the Schr\"{o}dinger equation in momentum space for a linear potential. It
should be noted that the accuracy of calculating the spectrum of the Schr\"{o}dinger
equation with a linear potential in the momentum space of both methods far exceeds the
accuracy of the solution in the approaches proposed in the papers
\cite{PhysRevD.88.076006,Deloff:2006xx,PhysRevD.47.3027,Leitao:2014jha,Tang:2001ii,Chen2013}.

\subsection{Special case for $\ell=0$}

By making use of mapping
\begin{equation}
\tilde{k}= \beta_{0}\sqrt{\frac{1+z}{1-z}}\; ,\; \; \;  \tilde{k}^{\prime}=
\beta_{0}\sqrt{\frac{1+t}{1-t}}\; ,\label{kt3j}
\end{equation}
we transform Eq.~ (\ref{eq1dd}) to
\begin{multline}
\frac{1}{\pi \hskip 1pt \beta_{0}}\hskip 1pt \left({\frac{1-z}{1+z}}\right)\int_{-1}^{1}
Q_{\ell}^{\prime}(y(t,z))\hskip 1pt
\frac{\phi_{n \ell}(t)\hskip 1pt \mathrm{d}t}{(1-t)\sqrt{1-t^2}}\\
=\left(\varepsilon_{n \ell} - \beta_{0}^{2}\left(\frac{1+z}{1-z}\right)\right)\hskip 1pt
\phi_{n \ell}(z) \;. \label{eq3d}
\end{multline}

In the case $\ell=0$, equation (\ref{eq3d}) after simplifications is written in the form
\begin{multline}
-\frac{1}{\pi \hskip 1pt \beta_{0}}\hskip 1pt \left(1-z\right)^{2}\int_{-1}^{1}\phi_{n
0}(t)\hskip 1pt
\sqrt{\frac{1+t}{1-t}}\hskip 1pt \frac{\mathrm{d}t}{(t-z)^{2}}\\
=\left(\varepsilon_{n 0}
- \beta_{0}^{2}\frac{1+z}{1-z}\right)\hskip 1pt \phi_{n 0}(z) \;. \label{eq4d}
\end{multline}

Starting from the structure of the integral equation, the most suitable interpolation
polynomial for the quadrature formula is the polynomial $V_n(t)$, and the weight function
can be chosen in the form
$$
w(t)=\sqrt{\frac{1+t}{1-t}}\; .
$$

As a result, the matrix of the eigenvalues problem takes the form
\begin{equation}
H_{i j}=  \left.\Big[\beta_{0}^{2}\hskip 2pt\delta_{i, j}\hskip 2pt \bar{k}_j -
\frac{\omega_{j}^{H} \left(\xi_{i,N}\right)}{\pi \hskip 1pt \beta_{0} \hskip 1pt
\overline{dk}_i} \right. \Big]\; ,\label{eq6d}
\end{equation}
where $\xi_{i,N}$ are the zeros of the polynomial $V_N(t)$ (see, Eq.~ (\ref{vnw})) and
the matrix elements $\omega_{j}^{H} \left(\xi_{i,N}\right)$ are calculated with the help
of Eq.~(\ref{vn-8}).

Therefore, it is possible to  compare the results of numerical calculations with matrix
(\ref{eq6d}) and exact values $-z_{n}$ (\ref{airy}). Table \ref{tab-3a} lists the values
of the relative error (\ref{deltotn})
\begin{equation}
\delta=\left|\frac{\varepsilon_{n 0}^{L}-\varepsilon_{n}^{(N)}}{\varepsilon_{n
0}^{L}}\right| \;, \label{deltotnzn}
\end{equation}
where $\varepsilon_n^{(N)}$ is the energy spectrum obtained by the numerical solution of
the eigenvalues problem for matrix (\ref{eq6d}) at the given number of $N$.
\begin{table}[h b t p]
\caption{Relative error $\delta_{n0}$ of solving Eq.~(\ref{eq4d}).}
\begin{ruledtabular}
\begin{tabular}{cccccc}
  $N$  & $n=1$ & $n=2$ & $n=3$ & $n=4$ & $n=5$ \\
  \hline
  50  & $3.4(-22)$   & $ 3.6(-20)$  & $ 3.2(-17)$  & $ 2.9(-15)$  & $ 8.0(-14)$\\
  100 & $1.7(-39)$   & $ 1.1(-35)$  & $ 1.5(-32)$  & $ 4.3(-31)$  & $ 5.2(-28)$\\
  150 & $ 4.5(-54)$  & $ 8.2(-50)$  & $ 4.8(-47)$  & $ 1.3(-43)$  & $ 5.9(-42)$\\
\end{tabular}
\end{ruledtabular}
\label{tab-3a}
\end{table}
Note, however, that the special method presented here  gives high-precision results only
in the case $\ell=0$. If $\ell \geqslant 1$, the kernel of Eq.~ (\ref{eq4d}) changes,
which leads to a sharp decrease of accuracy.

\section{Energy spectrum for Cornell potential}
\label{sect-7}

We are going to consider the case where both the Coulomb and the linear confinement
potential are present.  For the Cornell potential $V\left(r\right)=-\alpha/r+ \sigma
\hskip 1pt r$, there are no analytical solutions. Therefore, the numerical solution in
momentum space will be compared to the solution of this equation in coordinate space.

The method for estimating the accuracy of the solution will be the same as for the case
of a linear confinement potential.

\subsection{Quadrature scheme for $\ell \geqslant 0$}

From an analysis of the methods for solving the Schr\"{o}dinger equation in momentum
space for the Coulomb and linear potentials, the most optimal is the use of quadrature
formulas (\ref{qm1}) and (\ref{vn-12tn1}) in which the weight factors
$\omega_{i}^{HS}\left(z\right)$  (\ref{qm6}) and
$\omega_{i}^{\mathrm{log}}\left(z\right)$ (\ref{sp-7k}) depend on double-pole and
logarithmic singularities.

Using (\ref{pr1d}) and subtraction term (\ref{contr1}), the Schr\"{o}dinger equation with
Cornell potential $V\left(r\right)=-\alpha/r+\sigma\hskip 1pt r$ in momentum space
\begin{multline}
\frac{k^2}{2\mu}\phi_{\ell}(k)-\frac{\alpha}{\pi\hskip 1pt k} \int_{0}^{\infty}
Q_{\ell}(y)\hskip 1pt k^{\prime}\hskip 1pt \phi_{\ell}(k^{\prime})
\mathrm{d}k^{\prime}\\
+\frac{\sigma}{\pi\hskip 1pt k^{2}} \int_{0}^{\infty} Q_{\ell}^{\prime}(y)\hskip 1pt
\phi_{\ell}(k^{\prime}) \mathrm{d}k^{\prime}=E\phi_{\ell}(k) \; \label{eq1k}
\end{multline}
is written in the form
\begin{multline}
\left(\varepsilon_{n \ell} - \tilde{k}^2 \right) \phi_{n \ell}(\tilde{k})  \\
= \frac{1}{\pi \tilde{k}^2} \int_0^\infty  \left. \Big\{ Q^{\prime}_{\ell}(y) \phi_{n
\ell}(\tilde{k}^{\prime}) - Q^{\prime}_{0}(y)
\phi_{n\ell}(\tilde{k}) \right.\Big\}\mathrm{d} \tilde{k}^{\prime} \\
- \frac{\lambda}{\pi {\tilde{k}}} \int_0^\infty \,Q_\ell(y) \phi_{n
\ell}(\tilde{k}^{\prime}) \tilde{k}^{\prime}\hskip 1pt \mathrm{d} \tilde{k}^{\prime} \; ,
\label{cornel-1}
\end{multline}
where
\begin{equation}
\lambda=\cfrac{\alpha\hskip 1pt \left(2\mu\right)^{2/3}}{\sigma^{1/3}}\;. \label{lam1}
\end{equation}

The results of  calculation are presented in Table \ref{tab:cornel-1}.  To determine the
energy spectrum  the appropriate Schr\"{o}dinger equation was solved in both the momentum
and the coordinate space.  As seen from Table \ref{tab:cornel-1} there is excellent
agreement between these two methods of calculation.
\begin{table}[tb]
\caption{Relative errors $\delta_{n \ell}$ (\ref{deltotn}) on the computed Cornel binding
energies with $\lambda=1$.}
\begin{ruledtabular}
\begin{tabular}{llllll}
   &            &            &   $\ell=0$         &            &          \\
\hline
 {$N$} & \multicolumn{1}{c}{$n=1$} &
   \multicolumn{1}{c}{$n=2$} & \multicolumn{1}{c}{$n=3$} & \multicolumn{1}{c}{$n=4$} & \multicolumn{1}{c}{$n=5$} \\
\hline
 50     & $6.7(-13)$  & $ 2.4(-12)$  & $ 4.9(-12)$  & $ 1.9(-10)$  & $ 4.7(-9)$   \\
 100    & $2.7(-15)$  & $ 9.6(-15)$  & $ 2.0(-14)$  & $ 3.5(-14)$  & $ 5.2(-14)$  \\
 150    & $1.1(-16)$  & $ 3.7(-16)$  & $ 7.9(-16)$  & $ 1.4(-15)$  & $ 2.0(-15)$ \\
\hline
   &            &            &   $\ell=1$         &            &          \\
\hline
 {$N$} & \multicolumn{1}{c}{$n=1$} &
   \multicolumn{1}{c}{$n=2$} & \multicolumn{1}{c}{$n=3$} & \multicolumn{1}{c}{$n=4$} & \multicolumn{1}{c}{$n=5$} \\
\hline
 50     & $3.0(-12)$  & $ 1.4(-11)$  & $ 1.4(-10)$  & $ 4.7(-10)$  & $ 3.2(-8)$     \\
 100    & $3.2(-14)$  & $ 7.4(-14)$  & $ 1.1(-14)$  & $ 1.5(-13)$  & $ 3.1(-14)$    \\
 150    & $3.2(-15)$  & $ 6.0(-15)$  & $ 1.8(-15)$  & $ 1.2(-14)$  & $ 1.4(-15)$    \\
 \hline
   &            &            &   $\ell=2$         &            &          \\
\hline
 {$N$} & \multicolumn{1}{c}{$n=1$} &
   \multicolumn{1}{c}{$n=2$} & \multicolumn{1}{c}{$n=3$} & \multicolumn{1}{c}{$n=4$} & \multicolumn{1}{c}{$n=5$} \\ \hline
50     & $1.7(-10)$  & $ 1.8(-9)$   & $ 2.7(-9)$   & $ 1.3(-7)$   & $ 9.3(-7)$     \\
100    & $4.1(-16)$  & $ 1.1(-15)$  & $ 1.6(-15)$  & $ 3.2(-15)$  & $ 1.0(-13)$    \\
150    & $3.2(-17)$  & $ 7.2(-17)$  & $ 1.4(-16)$  & $ 2.2(-16)$  & $ 3.3(-16)$    \\
 \hline
   &            &            &   $\ell=3$         &            &          \\
\hline
 {$N$} & \multicolumn{1}{c}{$n=1$} &
   \multicolumn{1}{c}{$n=2$} & \multicolumn{1}{c}{$n=3$} & \multicolumn{1}{c}{$n=4$} & \multicolumn{1}{c}{$n=5$} \\ \hline
50    & $5.2(-10)$  & $ 4.5(-10)$  & $ 4.3(-8)$   & $ 3.0(-7)$   & $ 6.4(-7)$ \\
100   & $6.7(-20)$  & $ 1.1(-17)$  & $ 8.2(-16)$  & $ 8.2(-16)$  & $ 1.8(-13)$\\
150   & $1.7(-22)$  & $ 4.8(-22)$  & $ 1.1(-21)$  & $ 1.8(-21)$  & $ 1.7(-20)$\\
 \hline
   &            &            &   $\ell=4$         &            &          \\
\hline
 {$N$} & \multicolumn{1}{c}{$n=1$} &
   \multicolumn{1}{c}{$n=2$} & \multicolumn{1}{c}{$n=3$} & \multicolumn{1}{c}{$n=4$} & \multicolumn{1}{c}{$n=5$} \\ \hline
50     & $4.3(-10)$   & $ 4.0(-10)$  & $ 2.0(-8)$   & $ 9.1(-8)$   & $ 6.1(-7)$  \\
100    & $ 7.7(-17)$  & $ 1.7(-17)$  & $ 1.4(-14)$  & $ 1.2(-13)$  & $ 1.1(-13)$ \\
150    & $ 7.0(-24)$  & $ 2.8(-23)$  & $ 2.8(-21)$  & $ 2.3(-20)$  & $ 4.8(-20)$ \\
\end{tabular}
\end{ruledtabular}
\label{tab:cornel-1}
\end{table}

Thus, the use of the quadrature rules on the base of Eqs.~(\ref{qm1}) and
(\ref{vn-12tn1}) will allow us to find the spectrum of the system with the Cornell
potential with a relative error of $10^{-15}$ for $\ell=0$ and $10^{-22}$ for
$\ell>1$.

\subsection{Special quadrature scheme for $\ell = 0$}

Using (\ref{eq6v}) and (\ref{eq6d}) , matrix  $H_{i j}$ for the equation with Cornell
potential  at $\ell = 0$ is written in the form
\begin{multline}
H_{i j}=  \beta_{0}^{2}\hskip 1pt \delta_{i, j}\hskip 2pt \bar{k}_j -
\frac{\omega_{j}^{H} \left(\xi_{i,N}\right)}{\pi \hskip 1pt \beta_{0}\hskip 1pt
\overline{dk}_i} \\
-\frac{\lambda \hskip 1pt \beta_{0}}{\pi}\hskip 1pt
\dfrac{{\overline{dk}}_{j}}{\bar{k}_{i}^{1/2}}\hskip 1pt \omega_{j}^{Q_0}
\left(\xi_{i,N}\right)  \;,\label{eq6kkk}
\end{multline}

Table  \ref{tab-5a} represents the values
\begin{equation}
\delta_{n 0}=\left|\frac{\tilde{\varepsilon}_{n
0}-\varepsilon_{n}^{(N)}}{\tilde{\varepsilon}_{n 0}}\right| \; , \label{deltotn2}
\end{equation}
where $\tilde{\varepsilon}_{n 0}$ are the eigenvalues obtained in coordinate space.
\begin{table}[h b t p]
\caption{Value of $\delta_{n0}$ for polynomials $V_N(t)$ with $\lambda=1$.}
\begin{ruledtabular}
\begin{tabular}{cccccc}
  $N$  & $n=1$ & $n=2$ & $n=3$ & $n=4$  & $n=5$\\
  \hline
  50  & $4.9(-23)$   & $ 5.6(-20)$  & $ 1.7(-17)$  & $ 1.5(-15)$  & $ 9.7(-14)$    \\
  100 & $8.5(-40)$   & $ 2.7(-36)$  & $ 9.6(-34)$  & $ 1.9(-30)$  & $ 3.9(-28)$   \\
  150 & $5.0(-55)$   & $ 8.7(-51)$  & $ 2.8(-47)$  & $ 1.5(-44)$  & $ 1.4(-41)$   \\
\end{tabular}
\end{ruledtabular}
\label{tab-5a}
\end{table}

This method, as in the case of a special quadrature rules for a linear potential
(\ref{eq6d}), is highly accurate only for $\ell=0$. It is to be noted that numerical
calculations with the help of (\ref{eq6kkk}) completely agree with results
\cite{Kang:2006jd}.

\section{Conclusions}
\label{sect-8}

In this paper, we solve numerically the Schr\"{o}dinger equation in momentum space with
the Coulomb, linear confinement and Cornell potentials by using new quadrature rules.

The numerical results demonstrate the efficiency of the created method. The new
quadrature formulas, in which the singularities of integrals are included into the weight
functions, make it possible to solve the Schr\"{o}dinger equation for the momentum space
with high accuracy.

The achieved accuracy of calculations is many orders of magnitude higher than in similar
calculations in momentum space conducted in the papers
\cite{PhysRevD.88.076006,Deloff:2006xx,PhysRevD.47.3027,Leitao:2014jha,Tang:2001ii,Chen2013}.
Special high-precision methods of solution for states with zero orbital angular momentum
are considered.

These methods are easily generalized to the relativistic equations, where the potentials
are generally derived in momentum space. Consequently, the developed procedure to obtain
the energy spectra can be used to study and calculate various effects in the two-body
quantum systems, such as hydrogen-like atoms, hadronic atoms and bound quark systems.

\section{Acknowledgments}
\label{sect-9}

This work was supported by the Belarussian Foundation for Fundamental Research (Minsk,
Republic of Belarus).

The author is grateful to the Samara University (Samara, Russia) for technical support of
numerical calculations in the ``Wolfram Mathematica'' system.

\newpage
\appendix
\section{A subtraction term for the logarithmic singularity}
\label{pril-a}

We consider a one-dimensional integral,
\begin{equation}
\mathcal{C}_{\ell}=\int_{0}^{\infty} \frac{Q_{\ell}\left(y\right)}{k} \mathrm{d}k\; ,
\label{sc}
\end{equation}
where
\begin{equation}
y=\frac{k^2+p^2}{2\hskip 1pt  p\hskip 1pt  k} \; .\label{sc-1}
\end{equation}

Using the integral representation for Legendre polynomials of the second kind
$Q_{\ell}\left(y\right)$ \cite{Abramowitz1972}
\begin{equation}
Q_{\ell}\left(y\right)= \frac{1}{2}\int_{-1}^{1} \frac{P_{\ell}\left(x\right)}{y-x}
\mathrm{d}x \label{sc-2}
\end{equation}
we can modify  Eq.~~(\ref{sc}) as
\begin{equation}
\mathcal{C}_{\ell}=\int_{-1}^{1}P_{\ell}\left(x\right) \left[\lim\limits_{\eta \to 0}
\int_{0}^{\infty} \frac{p}{ (k^2+p^2+\eta^{2})-2\hskip 1pt  p \hskip 1pt k \hskip 1pt x }
\mathrm{d}k\right] \mathrm{d}x\; . \label{sc-3}
\end{equation}

After integrating over the variable $k$ in the Eq.~(\ref{sc-3}), we can obtain
\begin{equation}
\mathcal{C}_{\ell}=\int_{-1}^{1} P_{\ell}\left(x\right) \left[\frac{\pi}{2\hskip 1pt
\sqrt{1-x^2}}+\frac{\arcsin\hskip 1pt x}{ \sqrt{1-x^2}}\right] \mathrm{d}x \; .
\label{sc-4}
\end{equation}

Using the  integrals \cite{GradshteynE1980}
\begin{multline}
\frac{\pi}{2}\int_{-1}^{1} \frac{P_{\ell}\left(x\right)}{\hskip 1pt \sqrt{1-x^2}}
\mathrm{d}x \\
= \left\{\begin{array}{cl} \cfrac{\pi ^2}{2}\left[\cfrac{ (\ell-1)\hskip
1pt
!!}{\ell \hskip 1pt !!}\right]^{2}&\;, \; \ell =2\hskip 1pt m\\
 0&\;,\; \ell = 2\hskip 1pt m+1 \\
 \end{array}
 \right.\; ,\;  m=0,1,2\ldots
  \label{sc-5}
\end{multline}
and
\begin{multline}
\int_{-1}^{1} \frac{\arcsin\hskip 1pt x}{\hskip 1pt \sqrt{1-x^2}}\hskip 1pt
P_{\ell}\left(x\right) \mathrm{d}x \\
=\left\{\begin{array}{cl}
2\left[\cfrac{(\ell-1)\hskip 1pt !!}{\ell \hskip 1pt !!}\right]^{2}&\;,\ell =2\hskip 1pt m+1\\
 0  &\;,\ell = 2\hskip 1pt m   \\
 \end{array}
 \right.\; ,\;  m=0,1,2\ldots
  \label{sc-6}
\end{multline}
one obtains for the integral $\mathcal{C}_{\ell}$
\begin{multline}
\mathcal{C}_{\ell}=\int_{0}^{\infty} \frac{Q_{\ell}\left(y\right)}{k}
\mathrm{d}k=\left[\cfrac{(\ell-1)\hskip 1pt !!}{\ell \hskip 1pt !!}\right]^{2}\\
\times
\left\{\begin{array}{cl}
 \pi^2/2 &\;, \ell =2\hskip 1pt m \\
 2 &\;,\ell = 2\hskip 1pt m+1  \\
 \end{array}
 \right.\; ,\;  m=0,1,2\ldots
  \label{sc-7}
\end{multline}
where
$$
\ell \hskip 1pt !!=\ell(\ell-2)(\ell-4) \ldots\;, \ell \in \mathbb{Z}\; .
$$

The subtraction term (\ref{sc-7}) includes the known relation for the counter term with
$\ell=0$ (\ref{contrh1}) and it is somewhat easier to use in solving the equations than
the counter term (\ref{contrh2}) represented in \cite{PhysRevC.18.932}.

\section{The Chebyshev polynomials}
\label{pril-b}

Let us define the function
\begin{equation}
{K}_{n}^{\left(\alpha,\beta\right)}\left(z\right)=\left\{
\begin{array}{cll}
T_n(z)\; , &\alpha=~~\beta=-1/2 &\; ,  \\
U_n(z)\; , &\alpha=~~\beta=~~1/2  &\; ,  \\
V_n(z)\; , &\alpha=-\beta=-1/2 &\; ,   \\
W_n(z)\; , &\alpha=-\beta=~1/2  &\; ,   \\
\end{array}\right. \label{w5k}
\end{equation}
which includes the Chebyshev polynomials of  $1,2,3,4$ kinds (the details can be found in
\cite{Mason2002}).

The zeros of Chebyshev's polynomials are determined by the relation
\begin{equation}
\xi _{i,n}  = \cos\hskip 1pt \theta_{i,n}\; ,\; \;\; \;  \left(i = 1,\ldots, n \right) \;
. \label{f10s}
\end{equation}
The trigonometric representations and expressions for $\theta_{i,n}$ are given in Table
\ref{tabT-1}.
\begin{table}[h t p b]
\begin{ruledtabular}
\caption{Relations for Chebyshev polynomials ($x=\cos\theta$).}
\begin{tabular}{ccc}
Designation       & Expression   &  $\theta_{i,n}$       \\
\hline
$T_{n}\left(x\right)$   &$\cos n \theta$   & $\left(i-1/2\right)\pi/n$ \\
$U_{n}\left(x\right)$ &$\sin\left(n+1\right)\theta/\sin \theta$ & $i \hskip 1pt
\pi/(n+1)$\\
$V_n \left(x\right)$  &$\cos\left(n+1/2\right)\theta/\cos\left(\theta/2\right)$ &
$\left(2\hskip 1pt i-1\right)\hskip 1pt \pi/(2n+1)$  \\
 $W_n \left(x\right)$
&$\sin\left(n+1/2\right)\theta/\sin\left(\theta/2\right)$
& $\left(2\hskip 1pt i \right)\hskip 1pt \pi/(2n+1)$  \\
\end{tabular}
\end{ruledtabular}
\label{tabT-1}
\end{table}

We can prove that
\begin{eqnarray}
\frac{K_N^{\left(\alpha ,\beta\right)}\left(t\right)-K_N^{\left(\alpha
,\beta\right)}\left(z\right)}{t-z} \nonumber \\ = 2 \sum\limits_{j=0}^{N-1}
U_{N-1-k}(z)\hskip 1pt K_j^{\left(\alpha ,\beta\right)}\left(t\right) \nonumber
\\= 2 \sum\limits_{j=0}^{N-1} K_{N-1-j}^{\left(\alpha ,\beta\right)}(z)\hskip 1pt
U_j\left(t\right)\;. \label{w5}
\end{eqnarray}
In particular, $\alpha=\beta=-1/2$  for relation (\ref{w5}) leads to the expansion of the
form
\begin{multline}
\frac{T_N(t)-T_N(z)}{t-z} = 2 \hskip 1pt  \sideset{}{ \hskip 1pt ^{\prime}}
\sum\limits_{k=0}^{N-1} U_{N-1-k}(z)\hskip 1pt T_k(t)\\ = 2 \hskip 5pt  \sideset{}{
\hskip 1pt ^{\prime \prime}}\sum\limits_{k=0}^{N-1} T_{N-1-k}(z)\hskip 1pt U_k(t)\;,
\label{w6}
\end{multline}
where the signs $\prime$ and $\prime \prime$  indicate that the first and last term in
the sum is divided by two, respectively.

In order not to use notation with signs $\prime$ and $\prime \prime$  we introduce
additional functions
\begin{equation}
r_{k}^{n}=\left\{
\begin{array}{cll}
1/2\; ,  & k=n   & \; ,  \\
1  \; ,  & k\neq n   & \; ,
\end{array}\right. \label{af-1}
\end{equation}

\begin{equation}
c_{k}^{m}(n)=\left\{
\begin{array}{cll}
1/2\; ,  & k=n \mbox{~and}  & m \mbox{~is odd number} \; ,  \\
1  \; ,  & k=n \mbox{~and}  & m \mbox{~is even number}\; ,  \\
1  \; ,  & k\neq n  & \;.
\end{array}\right. \label{af-2}
\end{equation}

Using the integral and the equation (\ref{w5}),
\begin{equation}
\int_{-1}^1 U_{k}\left(t\right)\hskip 1pt \mathrm{d}t =
\frac{1-(-1)^{k+1}}{k+1}\label{uint}
\end{equation}
we calculate the analytic expression of the integral
\begin{equation}
R_{n}^{\left(\alpha,\beta\right)}\left(z\right) =\int_{ - 1}^1 \frac{K_n^{\left(\alpha
,\beta\right)}\left(t\right)-K_n^{\left(\alpha
,\beta\right)}\left(z\right)}{\left(t-z\right)}\hskip 1pt \mathrm{d}t \; .\label{w1}
\end{equation}

The integral (\ref{w1})
\begin{equation}
R_{n}^{\left(\alpha,\beta\right)}\left(z\right) = 4 \sum
_{i=0}^{\left[\frac{n-1}{2}\right]} \frac{K_{n-2i-1}^{(\alpha ,\beta )}(z)}{2 i+1}\; ,
\;  \left(\alpha,\beta = \pm 1/2\right) \label{w7}
\end{equation}
to the form for all cases, except in the case $\alpha=\beta=-1/2$ (Chebyshev polynomial
of the first kind). The $\left[n\right]$ symbol means that the integer part of the number
$n$ is taken.

If $\alpha=\beta=-1/ 2$, the Eq.~ (\ref{w7}) is transformed to the form
\begin{multline}
R_{n}^{\left(-1/2,-1/2\right)}\left(z\right)\\
 = 4 \sum _{i=0}^{b_n}
\frac{T_{n-2i-1}(z)}{2i+1}c_{i}^{n}(b_n) \; ,\; \; b_n=\left[\frac{n-1}{2}\right]\;
,\label{w7t}
\end{multline}
where $c_{k}^{m}(n)$ is defined by the equation  (\ref{af-2}). The presence of this
function leads to the fact that in the sum in (\ref{w7}) the last term is divided by two
if $n$ is an odd number.

\section{Matrix elements for the variational method}
\label{pril-3}

The most common method of numerical solving  Eq.~ (\ref{wfeq2}) with  centrally symmetric
potential $V\left(\left|\mathbf{r}\right|\right)$ is a variational method. In this
approach the solution of equation (\ref{wfeq2}) reduces to an eigenvalue problem
\begin{equation}\label{eqv2}
\sum^{\infty}_{k=0}\hskip 1pt a_{k} \left\langle
\Psi_{k}\right|\hat{H}\left|\Psi_{k^{\prime}}\right\rangle\equiv
\sum\limits^{\infty}_{k=0}\;\left\langle H\right\rangle_{k\;
k^{\prime}}a_{k^{\prime}}=E\; a_{k}
\end{equation}
by expanding the initial wave function  with respect to some complete set of trial wave
function $\Psi$
\begin{equation}\label{eqv3}
\Phi=\sum^{\infty}_{k=0}\;a_{k}\Psi_{k}\;.
\end{equation}
For an approximate solution of the series (\ref{eqv3}) terminate at a value $N-1$ and get
the eigenvalue problem
\begin{equation}\label{eqv4}
\sum^{N-1}_{k=0}\;a_{k} \left\langle
\Psi_{k}\right|\hat{H}\left|\Psi_{k^{\prime}}\right\rangle=\hat{E}_{n}\; a_{k^{\prime}}
\end{equation}
for the matrix $\left\langle H\right\rangle_{k\; k^{\prime}}$.

Entries of the matrix
$\left\langle\Psi_{n}\right|\hat{H}\left|\Psi_{n^{\prime}}\right\rangle=\left\langle
H\right\rangle_{n\; n^{\prime}}$, after calculating the angle part with the trial wave
functions
\begin{eqnarray}
&& \Psi_{n,\ell m}({\mathbf{r}}) =\psi_{n \ell}({r}){\cal Y}_{\ell
m}( \theta_{r},\phi_{r})\;, \nonumber \\
 && \widetilde{\Psi}_{n,\ell m}({\bf k})
=\widetilde{\psi}_{n\ell}({k}){\cal Y}_{\ell m}( \theta_{k},\phi_{k}) \label{norm}
\end{eqnarray}
represent the integrals of the form
\begin{eqnarray}
&&\left\langle H\right\rangle_{n\; n^{\prime}} = \int_{0}^{\infty}
\widetilde{\psi}^{*}_{n\hskip 1pt \ell} \left({{k}}\right) \left[\frac{{k}^2}{2
\mu}\right]\widetilde{\psi}_{n^{\prime}\hskip 1pt \ell} \left({{k}}\right) {{k}^{2}}
\mathrm{d}{k}+\nonumber \\
&&+ \int_{0}^{\infty} \psi^{*}_{n\hskip 1pt \ell} \left({{r}}\right)
V\left({r}\right)\psi_{n^{\prime}\hskip 1pt \ell} \left({{r}}\right)
{{r}^{2}}\mathrm{d}{r}\;. \label{eigenval}
\end{eqnarray}
The function $\widetilde{\psi}_{n\hskip 1pt  \ell} \left(k\right)$ is Fourier transform
of the wave function ${\psi}_{n \hskip 1pt \ell} \left({r}\right)$.

The principal aim here is to compute the expectation value of $V(\mathrm{r})={r}^{p}$
and  kinetic energy $k^2/(2\mu)$ with the trial pseudo-Coulomb wave functions
\begin{eqnarray}
\psi^{\mathrm{C}}_{n\ell} ({r}) &=& N_{n\ell}^{\mathrm{C}}\left(2\beta\right)^{3/2}\;
(2\beta \mathrm{r})^{\ell}e^{ - \beta
{r}} L_{n}^{2\ell + 2}(2\beta {r})\;, \nonumber \\
N_{n\hskip 1pt \ell}^{\mathrm{C}} &=& \sqrt{{\frac{{n!}}{{(n + 2\ell + 2)!}}}}\;,
\label{eqv7}
\end{eqnarray}
\begin{eqnarray}
\widetilde{\psi}^{\mathrm{C}}_{n\ell}(k)& = & \widetilde{N}_{n\hskip 1pt
\ell}^{\mathrm{C}}\left(\frac{\beta}{k^2+\beta^2}\right)^{\ell+2}\hskip 1pt k^{\ell}
\nonumber
\\
& \times &  P_{n}^{(\ell + 3/2,\ell+1/2)}(\frac{k^2-\beta^2}{k^2+\beta^2})\;, \nonumber
\\
\widetilde{N}_{n\ell}^{\mathrm{C}} &=& \frac{2\hskip 1pt \sqrt{{\beta\hskip 1pt
{n!}{(n+2\ell+2)!}}}}{\Gamma(n+\ell+3/2)}\; . \label{eqv8}
\end{eqnarray}
In Eqs.~(\ref{eqv7}) and (\ref{eqv8}) the functions  $L_{n}^{\ell}(z)$ and
$P_{n}^{\alpha,\beta}(z)$ are Laguerre and Jacobi polynomials accordingly. With these
basis functions the matrix element of the potential between the initial and final states
is calculated in configuration space.

The potential part of Eq.~(\ref{eigenval}) with $V({r})={r}^{p}$ written in the form
\begin{multline}
\langle{r^{p}}\rangle_{n \hskip 1pt  \ell, n^{\prime}} =\int_{0}^{\infty}
{\psi^{C}_{n \ell}(r,\beta)r^{p}\psi^{C}_{n^{\prime}\hskip 1pt \ell}(r,\beta){r}^{2}\mathrm{d}{r}}\\
=N^{C}_{n\hskip 1pt \ell}N^{C}_{n^{\prime}\hskip 1pt \ell}(2{\beta})^{2\ell+3} \\
\times \int_{0}^{\infty}{\mathrm{d}{r}\hspace{1pt} {r}^{2\ell+2+p}e^{-2{\beta}{r}}
L_{n}^{2\ell+2} {(2{\beta}{r})}L_{n^{\prime}}^{2\ell+2}{(2{\beta}{r})}} \;. \label{eqv15}
\end{multline}

A change of integration variable to the  combination $z=2{\beta}{r}$ in (\ref{eqv15})
yields
\begin{multline}
\langle{{r}^{p}}\rangle_{n \hskip 1pt  \ell, n^{\prime}}\\
=\frac{N^{C}_{n\hskip 1pt \ell}N^{C}_{n^{\prime}\hskip 1pt
\ell}}{(2\beta)^{p}}\int_{0}^{\infty}\mathrm{d}z\hspace{1pt} {z}^{2\ell+2+p}e^{-z}
L_{n}^{2\ell+2}(z)L_{n^{\prime}}^{2\ell+2}(z)\;. \label{eqv16}
\end{multline}
To use the Chu-Vandermonde sum formula with parameters $\alpha=2\ell+2$,
$\beta=2\ell+2+p$ \cite{GradshteynE1980}
\begin{equation}
L_{n-1}^{\alpha}(z)=\sum\limits^{n}_{j=1}\frac{(\alpha-\beta)_{n-j}}{(n-j)!}
L_{j-1}^{\beta}(z)\; , \label{cv}
\end{equation}
and the orthogonality relation for the Laguerre polynomials
\begin{equation}
\int_{0}^{\infty}dz{z}^{\beta}e^{-z}
L_{j}^{\beta}(z)L_{j^{\prime}}^{\beta}(z)=\frac{\Gamma(j+\beta+1)}{j!}\delta_{j,j^{\prime}}
\label{cv2}
\end{equation}
we obtain a general expression for the integral (\ref{eqv15})
\begin{multline}
 \langle \mathrm{r}^p \rangle_{n \ell,n^{\prime}}  =
\frac{1}{\left(2\beta\right)^{p}}\sqrt{\frac{n! \hskip 1pt  n^{\prime}!}{(n^{\prime} + 2
\ell+ 2)! (n + 2 \ell + 2)!}} \\
\times \sum\limits_{j=1}^{n+1} \hskip 1pt  \frac{\left(-p\right)_{n+1-j}
\left(-p\right)_{n^{\prime}+1-j}}{ \left(n+1-j\right)!\left(n^{\prime}+1-j\right)!}\\
\times \frac{\Gamma\left(2\ell+1+p+j\right)}{\left(j-1\right)!}\;,\;\;n \leqslant
n^{\prime}\;, \label{eqv17}
\end{multline}
where $(z)_{N}$ is Pochhammer symbol. Because of the symmetry of the matrix $\langle
\mathrm{r}^p \rangle_{n \ell,n^{\prime}}$ under interchange of $n^{\prime}$ and $n$ it is
easy to obtain the result for  $n > n^{\prime}$. Formula (\ref{eqv17}) generalizes the
relation  from \cite{Fulcher:1993sk}, where the calculations were made for special cases
with $p=-1$ and $p=1$.

From (\ref{eqv17}), we may deduce the  relations for $p=-1$ (Coulomb potential) and $p=1$
(linear potential) in the form
\begin{equation}\label{eqv18}
\langle{1/r}\rangle_{n \ell, n^{\prime}}
=\frac{\beta}{\ell+1}\sqrt{\frac{n^{\prime}!\hskip 1pt (n+2\ell +2)!}{ n!\hskip 1pt
(n^{\prime}+2 \ell+2)!}}\;,\;\;n \leqslant n^{\prime}\;,
\end{equation}
\begin{multline}
\label{eqv18v} \left\langle r \right\rangle_{n\ell, n^{\prime}}= \frac{1}{2 \beta}
\left.\bigg[  \delta_{n^{\prime},n}\hskip 1pt (2n+2\ell +3) \right.\\
-\delta_{n^{\prime},n-1}\hskip 1pt \sqrt{n \left(n+2\ell +2\right)}\\ \left.
-\delta_{n^{\prime},n+1}\hskip 1pt \sqrt{\left(n+1\right) \left(n+2\ell
+3\right)}\right.\bigg]\;,\;\;n \leqslant n^{\prime} \;.
\end{multline}

Part associated with the kinetic energy can be calculated exactly \cite{Fulcher:1993sk}
\begin{multline}
\label{eqv19} \left\langle k^2\right\rangle_{n\ell, n^{\prime}}=
\beta^{2} \sqrt{\frac{n^{\prime}!}{n!}}\sqrt{\frac{(n+2\ell +2)!}{(n^{\prime}+2\ell+2)!}} \\
\times\left(2+\frac{4n}{2 \ell+3}-\delta_{n^{\prime},\hskip 1pt n}\right)\; ,\;\;n
\leqslant n^{\prime} \;.
\end{multline}

\bibliographystyle{apsrev}
\bibliography{AndreevViktor}
\end{document}